\documentclass[         %
aps,                    
prd,                    
showpacs,               
superscriptaddress,    
nofootinbib,            
twocolumn,             %
showkeys,               %
preprintnumbers,        %
floatfix]               
{revtex4-1}               
\usepackage{graphicx,longtable,amsmath} 
\usepackage{mathrsfs,bm} 
\usepackage{mathtools} 
\usepackage{sublabel}

\newcommand{\msp}{\hspace{0.3pt}} 
\newcommand{\colvec}[2][.8]{%
  \scalebox{#1}{%
    \renewcommand{\arraystretch}{.8}%
    $\begin{pmatrix}#2\end{pmatrix}$%
  }
}
\begin{document} 
\title{ Neutrino Magnetic Moment, CP Violation and Flavor Oscillations in Matter} 
\author{Y. Pehlivan} 
\email{yamac.pehlivan@msgsu.edu.tr} 
\affiliation{Mimar Sinan Fine Arts University, Sisli, Istanbul, 34380, Turkey} 
\author{A. B. Balantekin} 
\email{baha@physics.wisc.edu} 
\affiliation{Department of Physics, University of Wisconsin - Madison, Wisconsin 53706 USA } 
\author{Toshitaka Kajino} 
\email{kajino@nao.ac.jp} 
\affiliation{National Astronomical Observatory of Japan 2-21-1 
Osawa, Mitaka, Tokyo, 181-8588, Japan} 
\affiliation{Department of Astronomy, University of Tokyo, Tokyo 113-0033, Japan} 
\date{\today} 
\begin{abstract} 
We consider collective oscillations of neutrinos, which are emergent
nonlinear flavor evolution phenomena instigated by neutrino-neutrino interactions
in astrophysical environments with sufficiently high neutrino densities.  We
investigate the symmetries of the problem in the full three flavor mixing scheme
and in the exact many-body formulation by including the effects of CP violation
and neutrino magnetic moment. We show that, similar to the two flavor scheme,
several dynamical symmetries exist for three flavors in the single-angle
approximation if the net electron background in the environment and the effects
of the neutrino magnetic moment are negligible.  Moreover, we show that these
dynamical symmetries are present even when the CP symmetry is violated in
neutrino oscillations. We explicitly write down the constants of motion through
which these dynamical symmetries manifest themselves  in terms of the generators
of the SU(3) flavor transformations.  We also show that the effects due to the
CP-violating Dirac phase factor out of the many-body evolution operator and
evolve independently of nonlinear flavor transformations if neutrino
electromagnetic interactions are ignored. In the presence of a strong magnetic
field, CP-violating effects can still be considered independently provided that
an effective definition for neutrino magnetic moment is used.  

\end{abstract} 
\medskip 
\preprint{} 
\maketitle 
 
\section{Introduction} 
\label{section: introduction} 

Among the known species neutrinos are the second most abundant particles in the
Universe after photons.  Many of them were created shortly after the Big Bang
and today form the Cosmic Neutrino Background the presence of which can be
inferred from the anisotropies of the Cosmic Microwave Background and the Cosmic
large-scale structure 
\cite{Lesgourgues:2012uu,Lesgourgues:2014zoa,Hannestad:2006zg,Dolgov:2002wy}.
Neutrinos also originate from various astrophysical sources such as core
collapse supernovae
\cite{Burrows:1990ts,Kotake:2005zn,Beacom:2010kk,Mathews:2014qba,Raffelt:1996} and black hole
accretion disks
\cite{Narayan:2001qi,Ruffert:1998qg,Popham:1998ab,Matteo:2002ck,Chen:2006rra,Malkus:2012ts}
where they are produced in copious amounts.  Given their abundance in the
Universe and the prevalence of extraordinary physical conditions in their
sources, it can be expected that even the tiniest anomalous electromagnetic
properties or CP violation features of neutrinos would have consequences in 
cosmology and astrophysics. 

Both in the Early Universe, and in the intense astrophysical sources mentioned
above, neutrinos are believed to undergo nonlinear forms of flavor evolution
which are generally termed \emph{collective neutrino oscillations}. These
oscillations follow from the self interactions of neutrinos which become
important when their number density is sufficiently high \cite{fuller_mayle} and
turn the flavor evolution into a many-body phenomenon
\cite{Pantaleone:1992eq,Pantaleone:1992xh,Sawyer:2005jk,Sigl:1992fn,Friedland:2006ke,Friedland:2003eh,Friedland:2003dv,Balantekin:2006tg,Duan:2010bg}.
The designation ``collective''  originates from the strong correlations that may
develop between neutrinos
\cite{Kostelecky:1994dt,Samuel:1995ri,Duan:2005cp,Duan:2006an,Duan:2010bg,Raffelt:2011yb}.


Although the collective oscillations of neutrinos are highly nonlinear, they
were shown to possess several dynamical symmetries under a set of idealized
conditions such as the absence of a net electron background, a two flavor mixing
scenario, and the so-called single angle approximation for the neutrino-neutrino
interactions \cite{Pehlivan:2011hp}.  These symmetries are dynamical in the
sense that the corresponding constants of motion depend on the interaction
parameters (unlike the space-time symmetries). It was also demonstrated that a
well known collective behavior of neutrinos, namely the spectral splits, is
connected to one of these symmetries \cite{Pehlivan:2011hp}.  One of the
purposes of this paper is to show that similar symmetries also exists for the
realistic three flavor mixing case and in the presence of a CP-violating Dirac
phase. 

The second purpose of this paper is to carefully examine the interplay between
the possible CP violation features of neutrinos, their anomalous electromagnetic
moments,  and the collective flavor oscillations with a particular focus on the
inherent many-body nature and the symmetries  of the latter. 

Our current understanding of particle physics already entails small anomalous
electric and magnetic dipole moments for neutrinos because the existence of the
neutrino mass calls for at least one right handed neutrino degree of freedom
which allows the neutrino to couple to photon at the one loop level
\cite{Lee:1977tib,Marciano:1977wx}. In the Standard Model, minimally extended to
include right-handed neutrinos, this yields very small values for the neutrino
dipole moments which are of the order of $10^{-19}\mu_B$ or smaller, where
$\mu_B$ denotes the Bohr magneton \cite{Balantekin:2013sda}.  However, various
theories beyond the Standard Model predict larger values.  The current
experimental upper limit on the anomalous magnetic moment of the neutrino is of
the order of $10^{-11} \mu_B$ \cite{Beda:2013mta} whereas a slightly better
upper limit of $10^{-12}\mu_B$ can be obtained from the constraints on
additional cooling mechanism for red giant stars due to plasmon decay into
neutrinos \cite{Raffelt:1999gv,Viaux:2013lha}.  For a recent review of the
electromagnetic properties of neutrinos, see Ref. \cite{Giunti:2014ixa}. 

Spin-flavor precession of neutrinos in magnetic fields was studied some time ago
\cite{Lim:1987tk}.  The effects of the neutrino magnetic moment on the
collective oscillations of neutrinos were recently examined in numerical
simulations \cite{deGouvea:2012hg,deGouvea:2013zp} and it was shown that
neutrinos and antineutrinos can swap their energy spectra if they propagate
through a strong magnetic field. Such a spectral swap can play an important role
in the $r$-process nucleosynthesis which could take place in the hot bubble
region of a core collapse supernova by transferring energy from the relatively
energetic antineutrinos of all flavors to the electron neutrinos and thereby
changing the electron fraction in the environment. 

On the other hand, the third neutrino mixing angle is shown to be nonzero by the
recent Daya Bay \cite{An:2013zwz}, RENO \cite{Ahn:2012nd} and Double Chooz
\cite{Abe:2012tg} experiments and this opens up the possibility of CP violation
in neutrino flavor oscillations. If the CP symmetry is broken by the neutrino
oscillations, the value of the corresponding Dirac phase may be within the reach
of the next generation very long base-line experiments such as LBNE
\cite{Adams:2013qkq} or LBNO \cite{::2013kaa}. 

The effects of a possible CP violation in supernova were considered by several
authors \cite{Balantekin:2007es,Kneller:2009vd,Gava:2010kz} in connection with
the collective oscillations. In particular, it was shown that, in the mean field
approximation, the term which contains the CP-violating phase factors out of 
the evolution operator so that CP-violating effects evolve independently of 
the nonlinear flavor transformations \cite{Gava:2008rp}.  However, a mean field
treatment depicts an interacting many-body system only approximately in terms of
independent particles moving in an average field which is collectively formed by
the particles themselves in a self consistent way.  Such a treatment by
definition ignores the quantum entanglements and takes into account only those
states in which each particle can be described in an effective one-particle
picture where the mean field consistency conditions can be met.  It is not clear
whether such a formulation allows us to easily distinguish the effects that are
induced by the dynamics from those that originate from a particular choice of
the initial conditions or from the reduction of the Hilbert space to unentangled
states. 

In this paper we show that the factorization of the CP-violating effects from
the flavor evolution during collective oscillations is more general than it is
implied in its original derivation. We use a formulation of the CP violation
which is independent of the mean field techniques and relies only on the
symmetry principles. We therefore show that even in the regime where quantum
entanglements due to many-body effects may be important, the CP-violating
effects factor out of the full many-body evolution operator and evolve
independently of the nonlinear flavor transformations. However, we also show
that, when neutrino magnetic moment comes into play in the  presence of a strong
magnetic field, the CP factorization procedure requires us to define an
\emph{effective} magnetic moment. This effective magnetic moment includes the
CP-violating Dirac phase and is different for neutrinos and antineutrinos,
indicating that the effects due CP violation and magnetic moment are intertwined
in the neutrino flavor evolution. But, as we argue in Section \ref{section:
magnetic moment}, the formulation introduced in this paper allows us to factor
the CP-violating phase out of the entire Hamiltonian including the effects due
to vacuum oscillations, matter refraction, self interactions and the
electromagnetic properties of the neutrinos at the expense of using an effective
definition for neutrino magnetic moment which is a small term and can be treated
perturbatively to the first order in most cases.

This paper is organized as follows: In Section \ref{section: transformations},
we introduce an operator formalism to describe flavor mixing of neutrinos. This
operator formulation is somewhat different from the commonly used mixing matrix
formalism, but it is better suited for the full many-body description of the
problem and for an analysis of the its symmetries. In Section \ref{section:
hamiltonian}, we express the vacuum oscillations of neutrinos together with the
refracting effects including neutrino self interactions in this formalism, and
describe how the CP-violating Dirac phase can be factored out of the total
Hamiltonian and the evolution operator (we do not consider the neutrino magnetic
moment at this point).  In Section \ref{section: constants of motion}, we
examine the dynamical symmetries of the problem in the single angle
approximation by ignoring effects of a possible net electron background and
present the corresponding many-body constants of motion. Although the
formulation of the neutrino self interactions are carried out entirely in the
exact many-body picture in this paper, in Section \ref{subsection: mean field}
we briefly consider an effective one-particle approximation in the form of a
mean field formulation and show that the expectation values of the many-body
constants of motion remain invariant under the mean field evolution of the
system.  In Section \ref{section: magnetic moment}, we include the effects of
neutrino magnetic moment in the presence of a uniform magnetic field and show
that the factorization of the CP-violating Dirac phase out of the full flavor
evolution Hamiltonian can be carried out using an effective definition for
neutrino magnetic moment.  

\section{Flavor Transformations} 
\label{section: transformations} 

In this paper, we use $a_{ih}$ and $b_{ih}$ to denote
the annihilation operators for neutrinos and antineutrinos, respectively, in the
$i^{\mbox{\footnotesize th}}$ mass eigenstate with 
chirality $h$. We consider only the ultra relativistic case for which the
helicity and chirality are the same for neutrinos and opposite for
antineutrinos. In other words,  $a_{ih}$ annihilate neutrinos with
helicity $h$ and $b_{ih}$ annihilate antineutrinos with helicity
$-h$.  If one does not take account of neutrino magnetic moment, which can
cause chirality to change, then it is sufficient to consider only the left
handed particles, i.e., negative helicity neutrinos and the positive helicity
antineutrinos.  For this reason, we drop the helicity index from our notation
and use 
\begin{equation} 
\label{convention}
a_i \equiv a_{i-} \quad\mbox{and}\quad b_i \equiv -b_{i-} 
\end{equation}  
until Section \ref{section: magnetic moment} where we take the neutrino magnetic
moment into account.  

In the literature, an isospin type formalism is typically employed in order to
describe a simplified two neutrino mixing scenario by introducing a
\emph{neutrino doublet} $(\nu_1, \nu_2)$ and the associated \emph{isospin
operators} (see, for example, Ref. \cite{Pehlivan:2011hp})
\begin{equation} 
\begin{split}
\label{isospin2} 
&J^+(\vec{p}\,)= a_{1}^{\dagger}(\vec{p}\,)a_{2}(\vec{p}\,)\msp,\qquad 
J^-(\vec{p}\,)= a_{2}^{\dagger}(\vec{p}\,)a_{1}(\vec{p}\,)\msp,\\ 
&\qquad J^z(\vec{p}\,)=\frac{1}{2}\left(a_{1}^{\dagger}(\vec{p}\,)a_{1}(\vec{p}\,)
-a_{2}^{\dagger}(\vec{p}\,)a_{2}(\vec{p}\,)\right)\msp, 
\end{split}
\end{equation} 
where $\vec{p}$ denotes the neutrino momentum. These operators form an SU(2) algebra. 

In the case of antineutrinos, the doublet $(-\bar{\nu}_2, \bar{\nu}_1)$ is
typically used instead of $(\bar{\nu}_1, \bar{\nu}_2)$ because it leads to
a unified treatment of neutrinos and antineutrinos and greatly simplifies the
formulation (see, for example, Refs.  \cite{Duan:2007fw, Duan:2007bt}). 
We can do so since under the SU(2) group the doublets $(-\bar{\nu}_2, \bar{\nu}_1)$ 
and $(\bar{\nu}_1, \bar{\nu}_2)$ transform with the same group element. 
Accordingly, the antineutrino isospin operators are defined as
\begin{equation} 
\begin{split}
\label{isospin2anti} 
&\bar{J}^+(\vec{p}\,)= -b_{2}^{\dagger}(\vec{p}\,)b_{1}(\vec{p}\,)\msp,\qquad 
\bar{J}^-(\vec{p}\,)= -b_{1}^{\dagger}(\vec{p}\,)b_{2}(\vec{p}\,)\msp,\\ 
&\qquad\bar{J}^z(\vec{p}\,)=\frac{1}{2}\left(b_{2}^{\dagger}(\vec{p}\,)b_{2}(\vec{p}\,)
-b_{1}^{\dagger}(\vec{p}\,)b_{1}(\vec{p}\,)\right)\msp. 
\end{split}
\end{equation}

The isospin formalism can be generalized to accommodate three generation mixing 
by introducing the following neutrino and antineutrino bilinears:
\begin{equation} 
\label{mass isospin operators} 
\begin{split}
T_{ij}(p,\vec{p}\,) &= a_i^\dagger (\vec{p}\,) a_j (\vec{p}\,)\msp,\\
T_{ij}(-p,\vec{p}\,) &=-b_j^\dagger (\vec{p}\,) b_i (\vec{p}\,)\msp, 
\end{split}
\end{equation}
for $i,j=1,2,3$. Here $p=|\vec{p}|$ denotes the energy of the neutrino.  Note
that we use the same notation for neutrino and antineutrino bilinears except
that the neutrino bilinears are labeled by the energy whereas the antineutrino bilinears
are labeled by minus the energy. Such a notation allows us to consolidate 
the neutrino and antineutrino degrees of freedom into one simple formulation in which
energy is allowed to run over both negative 
and positive values representing antineutrinos and neutrinos, respectively. In
order to do this, we introduce the convention 
\begin{equation}
T_{ij}(E,\vec{p}\,)
\mbox{ where }
\begin{cases}
E=p \quad\mbox{for neutrinos,}\\
E=-p \quad\mbox{for antineutrinos,}
\end{cases}
\end{equation}
and use the word \emph{energy} to refer to both positive and negative values
in the rest of this paper.  Let us also note here that we use the word
\emph{particle} generically to refer to both neutrinos and antineutrinos.

The operators defined in Eq. (\ref{mass isospin operators}) obey
$U(3)$ commutation relations\footnote{We refer to this group as $U(3)$ although it is technically the
tensor product of as many $U(3)$ algebras as the number of  particles in the
system. We use this offhand terminology throughout the paper for simplicity.}
\begin{equation}
\label{U3}
\begin{split}
&[T_{ij}(E,\vec{p}\,),T_{kl}(E^\prime\mkern-4mu,\vec{p}^{\,\prime})]=\\
&\mkern90mu\delta_{E,E^\prime}\delta_{\vec{p},\vec{p}^{\,\prime}} 
\left(\delta_{kj}T_{il}(E,\vec{p}\,)-\delta_{il}T_{kj}(E,\vec{p}\,)\right)\msp.
\end{split}
\end{equation}
In this equation, the factor $\delta_{\vec{p},\vec{p}^{\,\prime}}$ reflects the
fact that the particle operators corresponding to different momenta commute with one
another, whereas the factor $\delta_{E,E^\prime}$ guarantees that the neutrino and
antineutrino operators commute with each other even when they have the same
momentum. 

It is useful to introduce the sum\footnote{%
\label{continuum limit1}
In this paper we use sums over discrete momentum values rather than integrals
over the continuum values until Section \ref{section: magnetic moment} where we
switch back to the continuum integration (see footnote \ref{discrete} on page
\pageref{discrete} for the motivation behind this choice). 
Technically this requires the use of a
normalization volume $V$ such that every discrete sum over momentum is multiplied
by a factor of $1/V$ which yields 
\begin{equation*}
\frac{1}{V}\sum_{\vec{p}}\to\frac{1}{(2\pi)^3}\int d^3\vec{p}
\end{equation*}
in the continuum limit.  This introduces an overall $1/V$ factor multiplying
our Hamiltonian but we do not show this factor explicitly because 
the normalization volume becomes unimportant as soon as we take the continuum
limit in the sense that the physical quantities are independent of it.}
\begin{equation} 
\label{sumconvention} 
T_{ij}(E)\equiv
\sum_{\substack{\vec{p}\\(|\vec{p}|=p)}}T_{ij}(E,\vec{p}\,)\msp. 
\end{equation}
This sum runs over all neutrinos ($E=p$) or antineutrinos ($E=-p$) which travel
in different directions but have the same energy.  We would like to point out
that, since the collective oscillations of neutrinos are many-body phenomena,
one typically needs additional quantum numbers besides the momentum to label the
individual particles. However, we do not show these quantum numbers explicitly
in our formulas for ease of reading. 
Instead, when we use $\vec{p}$ as in Eq. (\ref{mass isospin operators}), for
example, we view it as a collective attribute which includes all the quantum
numbers needed to label an individual particle.  In any case, we consider these additional
quantum numbers to be also summed over in Eq.  (\ref{sumconvention}). 

We also introduce the sum over all particles of all energies
\begin{equation} 
\label{sumconvention2} 
T_{ij}\equiv\sum_{E} T_{ij}(E)\msp. 
\end{equation}
In this paper, a summation over energy such as the one in Eq.
(\ref{sumconvention2}), always runs over both positive and negative values so
that the resulting quantity incorporates both neutrinos and antineutrinos. Of course, 
we can always separate neutrino and antineutrino energy spectra when we need them. 

For three neutrino species, the transformation from mass to flavor basis can be
decomposed into three successive schemes of two-generation mixing.  For
this reason we first consider a transformation involving only the
$i^{\mbox{\footnotesize th}}$ and $j^{\mbox{\footnotesize th}}$ mass
eigenstates.  Note that the change from mass to flavor basis is a global
transformation in the sense that all neutrinos transform in the same way
irrespective of their energies. The same is also true for the antineutrinos
although neutrinos and antineutrinos transform differently in the presence of CP
violation. Such a transformation can be formulated in terms of the total
particle bilinears defined in Eq. (\ref{sumconvention2}). In particular, the
operators
\begin{equation}
\label{mass isospin ij} 
T_{ij}\msp, \quad 
T_{ji}\msp, \quad\mbox{and}\quad
\frac{1}{2}\left(T_{ii}-T_{jj}\right)\msp, 
\end{equation}
form an $SU(2)$ subalgebra%
\footnote{This algebra is $SU(2)$ rather than $U(2)$ because we did not include
the symmetric combination $T_{ii}+T_{jj}$.}
and generate the mixing between the 
$i^{\mbox{\footnotesize th}}$ and $j^{\mbox{\footnotesize th}}$ mass
eigenstates through the operator 
\begin{equation}
\label{Q and Qbar}
Q_{ij}(z)=e^{zT_{ij}}e^{\ln{(1+|z|^2)}\frac{1}{2}(T_{ii}-T_{jj})}e^{-z^*T_{ji}}\msp.
\end{equation}
Here $z$ is a complex variable which is related to the mixing angle
$\theta$ and a possible CP-violating phase $\delta$ by
\begin{equation}
\label{zij}
z=e^{-i\delta}\tan\theta.
\end{equation}
The operator in Eq. (\ref{Q and Qbar}) transforms the neutrinos as 
\begin{subequations}
\label{transformation ij}
\begin{equation}
\label{transformation ij a}
\begin{split}
     Q_{ij}^\dagger a_i(\vec{p}\,)Q_{ij}
     & = \cos\theta\: a_{i}(\vec{p}\,)+e^{-i\delta}\sin\theta\: a_{j}(\vec{p}\,)\msp, \\ 
     Q_{ij}^\dagger a_j(\vec{p}\,)Q_{ij}
     & = -e^{i\delta}\sin\theta\: a_{i}(\vec{p}\,)+\cos\theta\: a_{j}(\vec{p}\,)\msp,\\ 
\end{split}
\end{equation}
and the antineutrinos as 
\begin{equation}
\label{transformation ij b}
\begin{split}
     Q_{ij}^\dagger b_i(\vec{p}\,)Q_{ij} 
     & = \cos\theta\: b_{i}(\vec{p}\,)+e^{i\delta}\sin\theta\: b_{j}(\vec{p}\,)\msp, \\
     Q_{ij}^\dagger b_j(\vec{p}\,)Q_{ij}
     & = -e^{-i\delta}\sin\theta\: b_{i}(\vec{p}\,)+\cos\theta\: b_{j}(\vec{p}\,)\msp, 
\end{split}
\end{equation}
\end{subequations}
as can be easily shown by using the Baker-Champbell-Hausdorf formula 
\begin{equation}
e^A B e^{-A} = B + [A,B] + \frac{1}{2!} [A,[A,B]] + \dots  
\end{equation}
Note that, although neutrino and antineutrino bilinears appear symmetrically in
the definition of the operator $Q_{ij}$ (see Eqs. (\ref{sumconvention2}) and
(\ref{Q and Qbar})), 
the transformation of antineutrinos  differs from that of neutrinos in
Eq. (\ref{transformation ij}) by a complex phase in the presence of CP
violation, i.e., when $\delta\neq 0$. This is due to the difference in the
definitions of neutrino and antineutrino bilinears in Eq. (\ref{mass isospin
operators}). 

Mixing between three generations of neutrinos can be decomposed into three
consecutive transformations of two flavor mixing in the form of Eq. (\ref{transformation ij}). 
The relevant operator is
\begin{subequations}
\label{Q}
\begin{equation}
\label{Q decompose}
Q=Q_{23}(t_{\mbox{\tiny A}})Q_{13}(e^{-i\delta}t_{\mbox{\tiny R}})Q_{12}(t_{\odot})\msp,
\end{equation}
with
\begin{equation}
\label{zs}
t_{\odot}=\tan\theta_{\odot}\msp, \quad 
t_{\mbox{\tiny R}}=\tan\theta_{\mbox{\tiny R}}\msp, \quad 
t_{\mbox{\tiny A}}=\tan\theta_{\mbox{\tiny A}}\msp,  
\end{equation}
\end{subequations}
where $\theta_{\odot}$, $\theta_{\mbox{\tiny R}}$ and $\theta_{\mbox{\tiny A}}$
refer to solar, reactor and atmospheric mixing angles, respectively, and
$\delta$ is the CP-violating Dirac phase. 
With these definitions, the flavor and mass bases are simply related by
\begin{subequations}
\label{transformation}
\begin{equation}
\label{transformationa}
a_{\alpha_i}(\vec{p}\,) =  Q^\dagger a_i(\vec{p}\,)Q 
\quad \mbox{and} \quad
b_{\alpha_i}(\vec{p}\,) =  Q^\dagger b_i(\vec{p}\,)Q\msp, 
\end{equation}
where we set
\begin{equation}
\label{transformationb}
\alpha_1=e\msp, \quad \alpha_2=\mu\msp,\quad \alpha_3=\tau\msp. 
\end{equation}
\end{subequations}

In the literature, it is more common to express the relation between mass and
weak interaction bases with a mixing matrix rather than with an operator as in
Eq. (\ref{transformation}).
In fact, considering the successive two flavor transformations in Eq. 
(\ref{Q}) together with Eq. (\ref{transformation ij}) one
can write Eq. (\ref{transformation}) in the familiar form as  
\begin{subequations}
\label{mixing}
\begin{equation}
\label{transformation matrix}
\begin{pmatrix}
a_e \\ a_\mu \\ a_\tau
\end{pmatrix}
=W
\begin{pmatrix}
a_1 \\ a_2 \\ a_3
\end{pmatrix}
\qquad
\begin{pmatrix}
b_e \\ b_\mu \\ b_\tau
\end{pmatrix}
=W^*
\begin{pmatrix}
b_1 \\ b_2 \\ b_3
\end{pmatrix}\msp,
\end{equation}
where $W$ is a unitary matrix given by 
\begin{equation}
\label{mixing matrix}
W=
\colvec[0.80]{
     1 & 0 & 0  \\
     0 &  c_{23}  & s_{23} \\
     0 & - s_{23} &  c_{23} 
}
\colvec[0.80]{
     c_{13} & 0 &  s_{13}e^{-i\delta} \\
     0 &  1 & 0 \\
     - s_{13}e^{i\delta} & 0&  c_{13} 
}
\colvec[0.80]{
     c_{12} & s_{12} &0 \\
     - s_{12} & c_{12} & 0 \\
     0 & 0&  1 
}\msp,
\end{equation}
\end{subequations}
with $c_{ij}=\cos\theta_{ij}$ and $s_{ij}=\sin\theta_{ij}$.  But the operator
form of the neutrino mixing introduced in Eq.  (\ref{transformation}) is more
suitable for our purpose of formulating the many-body dynamics. 

The particle bilinears defined in Eq. (\ref{mass isospin operators})
can also be transformed into flavor basis using Eq. (\ref{transformation}),
e.g., 
\begin{equation}
\label{flavor isospin operators}
\begin{split}
T_{\alpha_i\alpha_j}(E,\vec{p}\,) &\equiv Q^\dagger T_{ij}(E,\vec{p}\,) Q\\
&=\begin{cases}
a_{\alpha_i}^\dagger (\vec{p}\,) a_{\alpha_j} (\vec{p}\,)\msp, \mbox{ for } E>0\msp,\\
-b_{\alpha_j}^\dagger (\vec{p}\,) b_{\alpha_i} (\vec{p}\,)\msp, \mbox{ for } E<0\msp.
\end{cases}
\end{split}
\end{equation}
These operators are subject to summation conventions which are analogous to those
introduced in Eqs. (\ref{sumconvention}) and (\ref{sumconvention2}). Note that
the transformation operator $Q$ has exactly the same form in both flavor and
mass bases. This can be shown as follows:
\begin{equation}
\label{Q in flavor}
Q=Q^\dagger Q Q=Q^\dagger Q_{23}Q_{13}Q_{12}Q=Q_{\mu\tau}Q_{e\tau}Q_{e\mu}\msp.
\end{equation}
Here $Q_{\alpha_i\alpha_j}(z)$ has the same form as $Q_{ij}(z)$ given in 
Eq. (\ref{Q and Qbar}) except that $i$ and $j$ replaced by $\alpha_i$ and
$\alpha_j$, respectively.

\section{Flavor Evolution of Neutrinos}
\label{section: hamiltonian}

\subsection{Vacuum Oscillations}
\label{subsection: Vacuum Oscillations}

Propagation of neutrinos and antineutrinos in vacuum is described by the
Hamiltonian 
\begin{equation}
\label{Hvacuum compact}
H_{\mbox{\footnotesize v}}=\sum_{\vec{p}}\sum_{i=1}^3\sqrt{p^2+m_i^2}
\big(T_{ii}(p,\vec{p}\,) - T_{ii}(-p,\vec{p}\,)\big)\msp. 
\end{equation}
Here $T_{ii}(E,\vec{p}\,)$ is a number operator in mass basis and  
is clearly conserved by the vacuum Hamiltonian, i.e., 
\begin{equation}
\label{conserved limit}
[H_{\mbox{\footnotesize v}}, T_{ii}(E,\vec{p}\,)]=0\msp.
\end{equation}
But since the neutrinos and antineutrinos are created in flavor states, the
initial state is not an eigenstate of the number operators in mass
basis. As a
result, although $T_{ii}(E,\vec{p}\,)$ is a constant of motion, it 
is not proportional to identity and cannot be subtracted from the Hamiltonian.
However the sum of the number operators over three generations 
has the same value in both the mass and flavor bases because of the unitarity of
the transformation. In other words, the initial state is an eigenstate of the
total number operator 
\begin{equation}
\label{total number}
\sum_{i=1}^3 T_{ii}(E,\vec{p}\,)
=\sum_{i=1}^3 T_{\alpha_i\alpha_i}(E,\vec{p}\,)\msp.
\end{equation}
Therefore the operator in Eq. (\ref{total number})  
is both constant \emph{and} proportional to identity 
which tells us that any multiple of it can be subtracted from the Hamiltonian. 
In particular, applying the ultra-relativistic approximation, 
\begin{equation}
\sqrt{p^2+m_i^2}\cong p+\frac{m_i^2}{2p}\msp,
\end{equation}
and subtracting the quantity
\begin{equation}
\label{subtract}
\sum_{E}\left[\left(E+\frac{m_1^2+m_2^2+m_3^2}{6E}\right)\sum_{i=1}^3T_{ii}(E)\right]
\end{equation}
allows us to express the Hamiltonian in Eq. (\ref{Hvacuum compact}) in terms of the
squared mass differences which are the relevant parameters for neutrino
oscillations. This yields 
\begin{equation}
\label{Hvacuum}
H_{\mbox{\footnotesize v}}=\sum_{E}\sum_{i=1}^3 \frac{\Delta_i^2}{6E}T_{ii}(E)\msp,
\end{equation}
where we defined
\begin{equation}
\label{Delta}
\Delta_i^2=\sum_{j (\neq i)}\delta m_{ij}^2\msp,
\end{equation}
and used the summation convention introduced in Eqs. (\ref{sumconvention}) and
(\ref{sumconvention2}).  As noted earlier, the sum over $E$ in Eqs.
(\ref{subtract}) and (\ref{Hvacuum}) runs over both neutrino ($E>0$) and
antineutrino ($E<0$) degrees of freedom. 

The vacuum Hamiltonian given in Eq. (\ref{Hvacuum}) can be expressed in flavor
basis by inverting Eq. (\ref{flavor isospin operators}), i.e., 
\begin{equation}
\label{vacuum Hamiltonian in flavor}
H_{\mbox{\footnotesize v}}=
\sum_{E}\sum_i \frac{\Delta_i^2}{6E} Q\;T_{\alpha_i\alpha_i}(E)\;Q^\dagger\msp.
\end{equation}
Here, all the information about the mixing angles and the CP-violating Dirac
phase is hidden in the operator $Q$.  If one applies the transformation imposed
by $Q$ using Eqs.  (\ref{transformation ij}), (\ref{flavor isospin operators})
and (\ref{Q in flavor}), then flavor off diagonal terms in the form of
$T_{\alpha_i\alpha_j}(p)$ appear in the Hamiltonian in Eq. (\ref{vacuum
Hamiltonian in flavor}) together with the mixing parameters. 

\subsection{Coherent Scattering in an Ordinary Background}
\label{subsection: Coherent Scattering of Neutrinos in Matter}

Neutrinos interact very weakly with matter but their flavor transformations are
nevertheless modified as they propagate in matter because their diminutive
scattering amplitudes primarily superpose coherently in the forward direction. As a
result, the dispersion relation is changed for each neutrino flavor depending on
its interactions in a way which is very similar to the refraction of light in a
medium \cite{Wolfenstein:1977ue,Mikheev:1986wj,Mikheev:1986gs}.  Although all
neutrinos undergo refraction, flavor oscillations are only sensitive to how
different flavors are distinguished from each other as they propagate.  In this
paper, we consider an ordinary matter background, i.e., a neutral and
unpolarized background composed of protons, neutrons, electrons and positrons.
Such an environment singles out electron neutrinos which experience an
additional charged current interaction. Therefore the net matter refraction effect is
captured by the Hamiltonian
\begin{equation}
\label{Hmatter}
H_{\mbox{\footnotesize m}}=\sqrt{2} G_F \mathcal{N}_e T_{ee}\msp.
\end{equation}
Here $\mathcal{N}_e$ denotes the net number density of electrons, (electrons minus
positrons) in the background and $T_{ee}$ is the total number of electron
neutrinos minus the total number of electron antineutrinos, i.e., 
\begin{equation}
T_{ee}=\sum_{\vec{p}}\left(
a_e^\dagger (\vec{p}\,) a_e (\vec{p}\,) -b_e^\dagger (\vec{p}\,) b_e (\vec{p}\,) \right)\msp,
\end{equation}
as implied by Eqs. (\ref{mass isospin operators}), (\ref{sumconvention}) 
and (\ref{sumconvention2}).  

\subsection{Self Interactions of Neutrinos}
\label{subsection: self interactions}

For sufficiently high neutrino densities, neutrino-neutrino scatterings can 
contribute to flavor evolution by creating a self refraction effect
\cite{fuller_mayle}. In the case of self interactions, it is not only the
forward scattering diagrams that add up coherently but also those  
diagrams in which particles exchange their flavors \cite{Pantaleone:1992eq}.  The
contribution of self interactions to the neutrino flavor evolution can be
described by the following effective Hamiltonian \cite{Sawyer:2005jk}: 
\begin{equation}
\label{Hself}
H_{\mbox{\footnotesize s}}=
\frac{G_F}{\sqrt{2}V} 
\sum_{i,j=1}^3
\mkern-2mu
\sum_{E,\vec{p}} 
\mkern-2mu
\sum_{E^\prime\mkern-4mu,\vec{p}^{\,\prime}}
\mkern-5mu
R_{\vec{p}\vec{p}^{\,\prime}}
T_{\alpha_i\alpha_j}(E,\vec{p}\,)
T_{\alpha_j\alpha_i}(E^\prime\mkern-4mu,\vec{p}^{\,\prime})\msp.
\end{equation}
Here 
\begin{equation}
\label{R}
R_{\vec{p}\vec{p}^{\,\prime}}=1-\cos\theta_{\vec{p}\vec{p}^{\,\prime}}\msp,
\end{equation}
where $\theta_{\vec{p}\vec{p}^{\,\prime}}$ is the angle between the momenta of the
interacting neutrinos and $V$ is the quantization volume\footnote{%
\label{continuum limit2}
We remarked earlier that we do not show the normalization volumes because they
are \emph{physically} not relevant, (see footnote \ref{continuum limit1} on page
\pageref{continuum limit1}). In
the case of neutrino self interactions, however, the normalization volume is
important because it determines the density of neutrinos which controls the
strength of the neutrino potential. Another way of saying this is that although,
for example, the vacuum oscillation term has only one $1/V$ factor, the self
interaction term has two such factors one of which tells us how many other
neutrinos our \emph{test neutrino} interacts with.}.

Self interactions turn neutrino flavor conversion into a many-body phenomenon
because the coherent superposition of flavor exchange diagrams couples the
flavor evolution of each neutrino to that of the entire ensemble.  This poses a
formidable problem because the resulting dynamics is non-linear and the presence
of the entangled states makes the dimension of the Hilbert space astronomically
large. The latter difficulty can be avoided by adopting an effective one particle
approximation which reduces the dimension of the Hilbert space by omitting
entangled many-body states. Such an approach was developed in Refs.
\cite{Pantaleone:1992eq,Pantaleone:1992xh} in the form of a mean field formalism
and is widely adopted by subsequent authors. However, the non-linearity of the
original many-body problem is inherited by the resulting mean field consistency
equations and renders them very difficult to solve in general.

Here, we do not necessarily resort to an effective one particle formulation but
neither do we attempt to solve the many-body problem. Our purpose is to examine
the full many-body system from the perspective of its symmetries in connection
with CP violation and dynamical invariants.  However, we also study the
manifestations of these symmetries under the effective one particle
approximation in Section \ref{subsection: mean field}.

Note that each term in the self interaction Hamiltonian given in Eq.
(\ref{Hself}) in the form of 
\begin{equation}
\sum_{i,j=1}^3T_{\alpha_i\alpha_j}(E,\vec{p}\,)T_{\alpha_j\alpha_i}(E^\prime\mkern-4mu,\vec{p}^{\,\prime})\msp,
\end{equation}
is a scalar in the flavor space, i.e., it is invariant under any global unitary
transformation. This follows from the fact that they all commute with the global
operators $T_{\alpha_k\alpha_l}$:
\begin{equation}
\label{rotational invariance}
[\sum_{i,j=1}^3T_{\alpha_i\alpha_j}(E,\vec{p}\,)T_{\alpha_j\alpha_i}(E^\prime\mkern-4mu,\vec{p}^{\,\prime})\;,\;
T_{\alpha_k\alpha_l}]=0\msp.
\end{equation}
Eq. (\ref{rotational invariance}), together with Eqs. (\ref{Q and Qbar}) and
(\ref{Q in flavor}), tells us that the self interaction Hamiltonian itself is
rotationally invariant, i.e.,  
\begin{equation}
\label{self rotation symmetry ij}
[H_{\mbox{\footnotesize s}},  Q_{\alpha_i\alpha_j}]=0
\end{equation}
is satisfied for every $i,j=1,2,3$. As a result, it has the same form in both the mass and
flavor bases:
\begin{eqnarray}
\label{Hself matter}
H_{\mbox{\footnotesize s}}&=&QH_{\mbox{\footnotesize s}}Q^\dagger \\
&=&
\frac{G_F}{\sqrt{2}V} \sum_{i,j=1}^3
\sum_{E,\vec{p}} \sum_{E^\prime\mkern-4mu,\vec{p}^{\,\prime}}
R_{\vec{p}\vec{p}^{\,\prime}}
T_{ij}(E,\vec{p}\,) T_{ji}(E^\prime\mkern-4mu,\vec{p}^{\,\prime})\msp.
\nonumber
\end{eqnarray}

\subsection{Neutrino Propagation with CP Violation}
\label{subsection: Neutrino Propagation with CP Violation}

The full problem of neutrino flavor evolution in an astrophysical environment, including vacuum oscillations,
matter effects and self interactions, is represented by the sum of the
Hamiltonians given in Eqs. (\ref{vacuum Hamiltonian in flavor}), (\ref{Hmatter})
and (\ref{Hself}):  
\begin{equation}
H= H_{\mbox{\footnotesize v}}+H_{\mbox{\footnotesize m}}+H_{\mbox{\footnotesize s}}\msp.
\label{total hamiltonian}
\end{equation} 
Here, the only term which explicitly involves the CP-violating phase is the
vacuum oscillation term through the operator $Q$. The matter term 
$H_{\mbox{\footnotesize m}}$ includes the net electron number density and it
can introduce a CP violation due to the matter-antimatter asymmetry of 
the background. But the matter Hamiltonian itself does not explicitly depend on the
\emph{intrinsic} CP-violating Dirac phase. A similar statement is also true for
the self interaction Hamiltonian $H_{\mbox{\footnotesize s}}$, i.e., although it can
introduce a CP asymmetry if the initial neutrino and antineutrino backgrounds
are not the same, neutrino-neutrino interactions are independent of the
intrinsic CP-violating phase. This can be seen from the fact that $H_{\mbox{\footnotesize s}}$ has the
same form in both the matter and flavor bases as indicated by Eqs. (\ref{Hself})
and (\ref{Hself matter}).

Factorization of the CP-violating phase from the flavor evolution stems
from the following identity which is true for any unitary operator in the form
of Eq. (\ref{Q and Qbar}):
\begin{subequations}
\label{identity}
\begin{equation}
\label{identity a}
Q_{e\tau}(e^{-i\delta}t_{\mbox{\tiny R}})=S_\tau^\dagger
Q_{e\tau}(t_{\mbox{\tiny R}}) S_\tau\msp.
\end{equation}
Here the operator $Q_{e\tau}(t_{\mbox{\tiny R}})$ 
does not contain the CP-violating Dirac phase $\delta$ which is now incorporated
into the operator  
\begin{equation}
\label{define S}
S_\tau=e^{-i\delta(T_{\tau\tau}+\bar{T}_{\tau\tau})}\msp.
\end{equation}
\end{subequations}
Therefore, we can write the transformation operator $Q$ in Eq. (\ref{Q in flavor}) as
\begin{equation}
\label{Q with S}
Q=Q_{\mu\tau}(t_{\mbox{\tiny A}})S_\tau^\dagger Q_{e\tau}(t_{\mbox{\tiny
R}})Q_{e\mu}(t_{\odot})S_\tau\msp,
\end{equation}
where we used Eq. (\ref{identity}) together with the fact that $S_\tau$ and
$Q_{e\mu}$ commute with each other because they live in orthogonal flavor
subspaces. However, $S_\tau$ does not commute with $Q_{\mu\tau}$ and for this
reason CP factorization cannot be realized in the ordinary flavor basis.
Instead, one has to transform into another basis in which $\mu$ and $\tau$
eigenstates are suitably mixed with one another. This is usually referred to as
the \emph{rotated flavor basis} and is defined as 
\begin{subequations}
\label{transformation mu-tau}
\begin{equation}
\label{rotated basis definition}
\begin{split}
&a_{\tilde{\alpha_i}}(\vec{p}\,)  \equiv Q_{\mu\tau}
a_{\alpha_i}(\vec{p}\,)Q_{\mu\tau}^\dagger\msp,
\\
&b_{\tilde{\alpha_i}}(\vec{p}\,)  \equiv Q_{\mu\tau}
b_{\alpha_i}(\vec{p}\,)Q_{\mu\tau}^\dagger\msp.
\end{split}
\end{equation}
From Eq. (\ref{transformation ij}), we see that this specifically
yields\footnote{Although $\nu_e$ and $\bar{\nu}_e$  remain the same under this
transformation which takes place in the orthogonal subspace, we introduce the
notation $\nu_{\tilde{e}}=\nu_e$ and $\bar{\nu}_{\tilde{e}}=\bar{\nu}_e$ because
it simplifies our formulas in subsequent sections.} 
\begin{equation}
\begin{aligned}
a_{\tilde{e}}(\vec{p}\,) & = a_{e}(\vec{p}\,)\msp,\\ 
a_{\tilde{\mu}}(\vec{p}\,) & 
=  \cos\theta_{\mbox{\tiny A}}\: a_{\mu}(\vec{p}\,)+\sin\theta_{\mbox{\tiny A}}\: a_{\tau}(\vec{p}\,)\msp, \\ 
a_{\tilde{\tau}}(\vec{p}\,) & 
=  -\sin\theta_{\mbox{\tiny A}}\: a_{\mu}(\vec{p}\,)+\cos\theta_{\mbox{\tiny
A}}\: a_{\tau}(\vec{p}\,)\msp, 
\end{aligned}
\end{equation}
for the neutrinos and
\begin{equation}
\begin{aligned}
b_{\tilde{e}}(\vec{p}\,) & = b_{e}(\vec{p}\,)\msp,\\ 
b_{\tilde{\mu}}(\vec{p}\,) & 
=  \cos\theta_{\mbox{\tiny A}}\: b_{\mu}(\vec{p}\,)+\sin\theta_{\mbox{\tiny A}}\: b_{\tau}(\vec{p}\,)\msp, \\ 
b_{\tilde{\tau}}(\vec{p}\,) & 
=  -\sin\theta_{\mbox{\tiny A}}\: b_{\mu}(\vec{p}\,)+\cos\theta_{\mbox{\tiny A}}\: b_{\tau}(\vec{p}\,)\msp, 
\end{aligned}
\end{equation}
\end{subequations}
for the antineutrinos. In most cases, the rotated and ordinary flavor
bases are physically equivalent to each other.  For example, in the case of
neutrinos emanating from  a supernova, $\nu_\mu$, $\nu_\tau$, $\bar{\nu}_\mu$
and $\bar{\nu}_\tau$ spectra are almost identical.  These neutrinos also undergo
the same neutral current weak interactions as they propagate in the mantle. As a
result, one has the same set of initial conditions and the same dispersion
relation in both the rotated and the ordinary flavor bases. 

That the desired factorization of CP-violating phase is achieved in the rotated
flavor base can be seen by multiplying Eq. (\ref{Q with S}) on the right with
$Q_{\mu\tau}^\dagger Q_{\mu\tau}$ and using Eq. (\ref{transformation mu-tau}).
The result is as follows:
\begin{equation}
\label{Q tilde}
Q =S_{\tilde{\tau}}^\dagger Q_{\tilde{e}\tilde{\tau}}(t_{\mbox{\tiny R}})
Q_{\tilde{e}\tilde{\mu}}(t_{\odot})S_{\tilde{\tau}}Q_{\mu\tau}(t_{\mbox{\tiny A}})\msp.
\end{equation}
Here, all operators with tilde signs have the same form as they are originally
defined except that $a_{\alpha_i}$ and $b_{\alpha_i}$ are replaced by
$a_{\tilde{\alpha_i}}$ and $b_{\tilde{\alpha_i}}$, respectively. In Eq. (\ref{Q
tilde}), the part of the transformation operator $Q$ excluding the rightmost
$Q_{\mu\tau}$ is now expressed in the rotated flavor basis and properly
factorized so as to separate the CP-violating phase from the flavor evolution.
The function of the rightmost $Q_{\mu\tau}$ is to transform the object on which
$Q$ is acting into the rotated flavor basis where the factorization is realized.
For example, using Eq. (\ref{Q tilde}), we can express the vacuum Hamiltonian
given in Eq.  (\ref{vacuum Hamiltonian in flavor}) as 
\begin{equation}
\label{hamiltonian with S inside}
\begin{split}
H_{\mbox{\footnotesize v}}&=
S_{\tilde{\tau}}^\dagger Q_{\tilde{e}\tilde{\tau}}(t_{\mbox{\tiny R}})
Q_{\tilde{e}\tilde{\mu}}(t_\odot)S_{\tilde{\tau}}
\\
&\mkern20mu
\times \sum_{E}\sum_i \frac{\Delta_i^2}{6E} 
T_{\tilde{\alpha}_i\tilde{\alpha}_i}(E)
S_{\tilde{\tau}}^\dagger
Q_{\tilde{e}\tilde{\mu}}^\dagger(t_\odot)
Q_{\tilde{e}\tilde{\tau}}^{\dagger}(t_{\mbox{\tiny R}}) S_{\tilde{\tau}}\msp,
\end{split}
\end{equation}
where we applied the definition of the rotated flavor basis from Eq.
(\ref{transformation mu-tau}) in order to transform $ T_{\alpha_i\alpha_i}(E)$
to $T_{\tilde{\alpha}_i\tilde{\alpha}_i}(E)$. One should also note that
$S_{\tilde{\tau}}$ commutes with $T_{\tilde{\alpha}_i\tilde{\alpha}_i}(E)$
because $S_{\tilde{\tau}}$ involves only the number operators in the rotated
flavor basis and $T_{\tilde{\alpha}_i\tilde{\alpha}_i}(E)$ are also number
operators themselves. Applying this to Eq. (\ref{hamiltonian with S inside})
leads to the result
\begin{equation}
\label{Hvacuum factorized}
H_{\mbox{\footnotesize v}}=S_{\tilde{\tau}}^\dagger
H_{\tilde{\mbox{\footnotesize v}}}^{(0)}S_{\tilde{\tau}}\msp,
\end{equation}
where $H_{\tilde{\mbox{\footnotesize v}}}^{(0)}$ is the Hamiltonian which represents the vacuum
oscillations in the rotated flavor basis in the absence of any CP-violating
phase. It is given by  
\begin{equation}
\begin{split}
H_{\tilde{\mbox{\footnotesize v}}}^{(0)}&=
Q_{\tilde{e}\tilde{\tau}}(t_{\mbox{\tiny R}})
Q_{\tilde{e}\tilde{\mu}}(t_\odot)\\
&\mkern20mu
\times \sum_{E}\sum_i \frac{\Delta_i^2}{6E} 
T_{\tilde{\alpha}_i\tilde{\alpha}_i}(E)
Q_{\tilde{e}\tilde{\mu}}^\dagger(t_\odot)
Q_{\tilde{e}\tilde{\tau}}^{\dagger}(t_{\mbox{\tiny R}})\msp.
\end{split}
\end{equation}
 
The matter Hamiltonian given in Eq. (\ref{Hmatter}) and the
self interaction Hamiltonian given in Eq. (\ref{Hself}) both commute with 
the CP violation term $S_{\tilde{\tau}}$: 
\begin{equation}
\label{commutators}
[H_{\mbox{\footnotesize m}},S_{\tilde{\tau}}]=0
\quad\mbox{and}\quad
[H_{\mbox{\footnotesize s}},S_{\tilde{\tau}}]=0\msp.
\end{equation}
The first commutator above is trivially true because $H_{\mbox{\footnotesize m}}$ and
$S_{\tilde{\tau}}$ live in orthogonal flavor spaces and the second commutator
immediately follows from Eq. (\ref{rotational invariance}) with $k=l$. At a more
intuitive level, the second commutator in Eq. (\ref{commutators}) is a result of  
the fact that the scattering of neutrinos from each other does not change the
total number of neutrinos or antineutrinos in any flavor eigenstate and that the
operator $S_{\tilde{\tau}}$ includes only the total number operators for
$\nu_{\tilde{\tau}}$ and $\bar{\nu}_{\tilde{\tau}}$ as can be seen from its
definition in Eq. (\ref{define S}). Therefore, the Hamiltonian
in Eq.  (\ref{total hamiltonian}) can be written as 
\begin{equation}
\label{total hamiltonian factorized}
H=S_{\tilde{\tau}}^\dagger \left(H_{\tilde{\mbox{\footnotesize v}}}^{(0)}
+H_{\mbox{\footnotesize m}}+H_{\mbox{\footnotesize s}}\right)S_{\tilde{\tau}}\msp.
\end{equation} 

The CP-violating phase is now factorized in such a way that the Hamiltonian
inside the parenthesis in Eq. (\ref{total hamiltonian factorized}) has no 
CP-violating phases. In this Hamiltonian the vacuum term is expressed in
the rotated flavor basis whereas the other terms are written in ordinary flavor
basis. However, both $H_{\mbox{\footnotesize m}}$ and $H_{\mbox{\footnotesize s}}$ do not change under the
transformation from ordinary to rotated flavor bases because they both commute
with the transformation operator $Q_{\mu\tau}$:
\begin{equation}
\label{self matter mu tau commutator}
[H_{\mbox{\footnotesize m}},Q_{\mu\tau}]=0
\quad\mbox{and}\quad
[H_{\mbox{\footnotesize s}},Q_{\mu\tau}]=0\msp.
\end{equation}
The first commutator in Eq. (\ref{self matter mu tau commutator}) is again
trivially true since $\nu_e$ is orthogonal to the $\nu_\mu$-$\nu_\tau$ subspace
and it leads to
\begin{subequations}
\label{self matter mu tau invariance}
\begin{eqnarray}
\label{matter mu tau invariance}
H_{\mbox{\footnotesize m}}=Q_{\mu\tau}H_{\mbox{\footnotesize m}}Q_{\mu\tau}^\dagger = H_{\tilde{\mbox{\footnotesize m}}}
=\sqrt{2} G_F \mathcal{N}_e T_{\tilde{e}\tilde{e}}\msp.
\end{eqnarray}
The second commutator in Eq. (\ref{self matter mu tau commutator})  is a special
case of Eq. (\ref{self rotation symmetry ij}) and it allows us to write 
\begin{eqnarray}
\label{self mu tau invariance}
H_{\mbox{\footnotesize s}} &=& Q_{\mu\tau}H_{\mbox{\footnotesize s}}Q_{\mu\tau}^\dagger =H_{\tilde{\mbox{\footnotesize s}}}\\ 
&=&\frac{G_F}{\sqrt{2}V} \sum_{i,j=1}^3
\sum_{E,\vec{p}}
\sum_{E^\prime\mkern-4mu,\vec{p}^{\,\prime}}
R_{\vec{p}\vec{p}^{\,\prime}}
T_{\tilde{\alpha}_i\tilde{\alpha}_j}(E,\vec{p}\,)
T_{\tilde{\alpha}_j\tilde{\alpha}_i}(E^\prime\mkern-4mu,\vec{p}^{\,\prime})\msp.
\nonumber
\end{eqnarray}
\end{subequations}
Therefore, the total Hamiltonian given in Eq. (\ref{total hamiltonian factorized}) 
can be written as 
\begin{subequations}
\label{total hamiltonian factorized compact}
\begin{equation}
\label{total hamiltonian factorized compact a}
H=S_{\tilde{\tau}}^\dagger
\tilde{H}^{(0)}S_{\tilde{\tau}}\msp,
\end{equation} 
where $\tilde{H}^{(0)}$ is an Hamiltonian which 
describes the vacuum oscillations and coherent
scatterings of neutrinos from the background particles as well as from each
other in the rotated flavor space and includes no CP-violating phase. 
It is given by 
\begin{equation}
\label{CP free H in rfb}
\tilde{H}^{(0)} 
= H_{\tilde{\mbox{\footnotesize v}}}^{(0)}+H_{\tilde{\mbox{\footnotesize m}}}+H_{\tilde{\mbox{\footnotesize s}}}\msp.
\end{equation} 
\end{subequations}
Eq. (\ref{total hamiltonian factorized compact}) tell us that the
collective flavor transformations of neutrinos \emph{as a many-body system} can
be described by an evolution operator 
\begin{subequations}
\label{U}
\begin{equation}
U(t)=S_{\tilde{\tau}}^\dagger \tilde{U}^{(0)}(t)S_{\tilde{\tau}}\msp,
\end{equation}
where $\tilde{U}_0(t)$ is the evolution operator corresponding to the
Hamiltonian $\tilde{H}^{(0)}$. In other words, it is the solution of  
\begin{equation}
i\hbar\frac{d}{dt}\tilde{U}^{(0)}(t)=\tilde{H}^{(0)}\tilde{U}^{(0)}(t)\msp,
\end{equation}
\end{subequations}
with the initial condition $\tilde{U}_0(t=0)=I$. 

\section{Constants of Motion}
\label{section: constants of motion}

Self interactions turn the problem of neutrino flavor transformation in an
astrophysical environment into a many-body phenomenon and give rise to highly
non-linear forms of flavor evolution. Still, numerical simulations reveal that
some forms of collective regular behaviour can emerge from the apparent
complexity. \emph{Synchronized oscillations} in which all neutrinos oscillate
with a single frequency \cite{Kostelecky:1994dt} and \emph{bi-polar
oscillations} in which the whole ensemble can be described in terms of two
frequencies \cite{Samuel:1995ri} are the earliest discoveries of such
collective behaviour and both were observed in a simplified two neutrino mixing scheme under
the mean field approximation.  Another noteworthy emergent behaviour is the
phenomenon of spectral splits in which neutrinos or antineutrinos exchange their
energy spectra at certain critical energies under the adiabatic evolution
conditions \cite{Duan:2006an}. These splits are observed in numerical
simulations for both two and three flavor mixing scenarios in the mean field
case. For a review, see Ref.  \cite{Duan:2010bg}.

Such collective modes of regular behavior call attention to possible symmetries
which may underline the dynamics of the system. In fact, an earlier study
\cite{Pehlivan:2011hp} by the present authors pointed out to some parallels
between self interacting neutrinos in a two flavor mixing scheme and the BCS
model of superconductivity \cite{Bardeen:1957mv} describing the Cooper pairs of
electrons in the conduction band of a metal. In particular, the role of the
neutrino flavor isospin (see Eqs. (\ref{isospin2}) and (\ref{isospin2anti})) in the
former case is played by the pair quasispin in the latter. We used this
analogy to show that certain dynamical symmetries, which were already known in
the context of the BCS model \cite{Gaudin1,Gaudin2,Cambiaggio}, are also
respected by flavor oscillations of self interacting neutrinos in the exact
many-body case if the following conditions are satisfied: 
\begin{enumerate}
\item The single angle approximation is adopted,
\item no net leptonic background is present, and 
\item the neutrino density is fixed. 
\end{enumerate}
There are as many such dynamical symmetries as the number of energy modes under
consideration and they manifest themselves as a set of \emph{constants of
motion}, i.e., quantities which depend non-trivially on the initial flavor
content of the ensemble and do not change as neutrinos propagate and undergo
flavor evolution.

It was also shown in Ref. \cite{Pehlivan:2011hp} that under the adoption of an
effective one particle approximation (i.e., in the mean field picture), these dynamical
symmetries are no longer exact but the expectation values of the corresponding constants of
motion continue to remain invariant. These mean field invariants are closely related to the
\emph{N-mode} coherence modes considered in Ref. \cite{Raffelt:2011yb} which are
also known as \emph{degenerate solutions} in the context of the BCS model
\cite{yuzb,yuzb2}. 

It should be noted that, although these dynamical symmetries are exact only
under the assumptions listed above, they can still be relevant when the system
is away from these idealized conditions. For example, it was demonstrated in
Ref.  \cite{Pehlivan:2011hp} that the two flavor spectral split phenomenon,
which emerges as neutrinos adiabatically evolve from a high density region into
the vacuum can be analytically understood in terms of one of the dynamical
symmetries although condition 3 is violated in this case. In this scheme, the
split frequency corresponds to the chemical potential in the BCS model of
superconductivity.

These observations clearly call for a thorough analysis of collective neutrino
oscillation modes in connection with the dynamical symmetries which will be
the subject of a future study. In this paper, we restrict ourselves solely to a 
study of the symmetries themselves. In particular we show that the dynamical
symmetries and the associated constants of motion, which were originally found
in the two flavor mixing scheme using an analogy to the BCS model, can be
generalized to the full three flavor mixing case.  We also show that these dynamical
symmetries continue to be exact even when the CP symmetry is broken by neutrino
oscillations.

\subsection{In the Exact Many Body Picture}

In the light of above comments, we ignore any net electron background in this section,
adopt the single angle approximation for neutrino self interactions and assume
that neutrinos occupy a fixed volume.  The single angle approximation assumes
that all neutrinos experience the same flavor transformation regardless of their
direction of travel, which amounts to replacing the angular factor
$R_{\vec{p}\vec{q}}$ introduced in Eq. (\ref{R}) with a suitable
representative value $R$. In this case, the Hamiltonian describing the flavor evolution of
neutrinos reduces to 
\begin{equation}
\label{vacuum and self hamiltonian}
H= \sum_{E}\sum_{i=1}^3 \frac{\Delta_i^2}{6E}T_{ii}(E)
+\frac{\mu}{2} \sum_{i,j=1}^3 T_{ij}T_{ji}\msp,
\end{equation}
where $\mu$ is given by
\begin{equation}
\mu=R\frac{\sqrt{2}G_F}{V}\msp.
\end{equation}
Here we used Eqs. (\ref{Hvacuum}) and (\ref{Hself matter}) 
in order to express the Hamiltonian in mass basis where
it takes a simpler form and we employed the 
summation conventions introduced in Eqs. (\ref{sumconvention}) and
(\ref{sumconvention2}). 

Using the $U(3)$ commutators given in Eq. (\ref{U3}), one can easily show that
the operators%
\footnote{\label{discrete}
In the continuum limit, the sum over $E^\prime$ is replaced by an integral which
has a singularity at $E^\prime=E$. But the integral does not diverge as will be
seen in Section \ref{subsection: mean field} below. In presenting the constants
of motion in this paper, we choose to use sum over discrete values of
energy-momentum because in practise one usually carries out the calculation over
discretized spectrum and we wanted to emphasize that in the discrete case,
$E^\prime=E$ term should be removed from the sum in Eq. (\ref{3 flavor
invariants mb}).} 
\begin{equation}
\label{3 flavor invariants mb}
h_E=\sum_{i=1}^3\frac{\Delta_i^2}{3} T_{ii}(E) 
+\mu\sum_{i,j=1}^3
\sum_{\substack{E^\prime\\ (E^\prime\neq E)}}
\frac{T_{ij}(E)T_{ji}(E^\prime)}{\frac{1}{2E}-\frac{1}{2E^\prime}}
\end{equation}
are constants of motion of the Hamiltonian given in Eq. (\ref{vacuum and self
hamiltonian}) because they commute with the Hamiltonian and with each
other, i.e., for every $E$ and $E^\prime$
\begin{equation}
\label{constants}
\left[{H},h_E\right]=0\quad\mbox{and}\quad
\left[h_E,h_{E^\prime}\right]=0
\end{equation}
are satisfied. Note that in Eqs. (\ref{3 flavor invariants mb}) and
(\ref{constants}), the energies $E$ and $E^\prime$ can take both positive and
negative values.  This tells us that for every \emph{physical energy} mode $p$
in the system, there are two constants of motion given by $h_{p}$ and $h_{-p}$
corresponding to neutrino and antineutrino degrees of freedom, respectively. The
Hamiltonian itself, which is given in Eq. (\ref{vacuum and self hamiltonian}),
can be written as a sum of these invariants, i.e., 
\begin{equation}
\label{Hamiltonian as sum}
H=\sum_E\frac{1}{2E}h_E\msp,
\end{equation}
up to some terms which are proportional to identity.

We would like to note that in the limit of $\mu \rightarrow 0$, 
self interactions of neutrinos disappear and the Hamiltonian given in Eq. (\ref{vacuum and self hamiltonian}) 
reduces to the vacuum propagation Hamiltonian only. In this limit, the
invariants presented in Eq. (\ref{3 flavor invariants mb}) reduce to number
operators for mass eigenstates and we recover Eq. (\ref{conserved limit}). 
However away from the $\mu \rightarrow 0$ limit, the invariants given in 
Eq. (\ref{3 flavor invariants mb}) are nontrivial and cannot be reduced to a combination of number operators. 
We also would like to note that both the Hamiltonian given in Eq. (\ref{vacuum
and self hamiltonian}) and the invariants given in Eq. (\ref{3 flavor invariants mb}) reduce to their two flavor
counterparts presented in Ref. \cite{Pehlivan:2011hp} if one restricts the sums
over three mass eigenstates to include only two of them (see Appendix \ref{Appendix
reduction}). 

One can express the constants of motion in the flavor basis using  
the inverse of Eq. (\ref{flavor isospin operators}) as 
\begin{eqnarray}
\label{3 flavor invariants fb}
h_E=Q\sum_{i=1}^3\frac{\Delta_i^2}{3} \mkern-20mu&&T_{\alpha_i\alpha_i}(E)Q^\dagger
\\
&+& \mu \sum_{i,j=1}^3 \sum_{E^\prime(\neq E)}
\frac{T_{\alpha_i\alpha_j}(E)T_{\alpha_j\alpha_i}(E^\prime)}{\frac{1}{2E}-\frac{1}{2E^\prime}}\msp,
\nonumber
\end{eqnarray}
where we also used Eq. (\ref{rotational invariance}) which tells us that the
quadratic part of the constants of motion will have the same form in both the
flavor and mass bases. 

It is important to note that Eq. (\ref{constants}) are valid even in the
presence of the CP-violating phase. In other words, the many-body dynamical
symmetries of the system are not broken when the neutrino oscillations are not
CP invariant. However, the CP-violating phase can be factored out in a way similar to
Eq. (\ref{total hamiltonian factorized compact}), i.e., 
\begin{equation}
\label{invariants factorized}
h_E=S_{\tilde{\tau}}^\dagger \tilde{h}_E^{(0)}S_{\tilde{\tau}}\msp,
\end{equation}
where $\tilde{h}_E^{(0)}$ are the constants of motion of the Hamiltonian 
\begin{equation}
\label{CP free H in rfb without matter}
\tilde{H}^{(0)} 
= H_{\tilde{\mbox{\footnotesize v}}}^{(0)}+H_{\tilde{\mbox{\footnotesize s}}}\msp,
\end{equation} 
which represents the vacuum oscillations and self interactions of neutrinos and
antineutrinos in the rotated flavor basis in the absence of any CP-violating
phase.  They are given by
\begin{eqnarray}
\label{3 flavor invariants rb}
h_E^{(0)}&=&
Q_{\tilde{e}\tilde{\tau}}(t_{\mbox{\tiny R}}) Q_{\tilde{e}\tilde{\mu}}(t_\odot)
\; \sum_{i=1}^3\frac{\Delta_i^2}{3} T_{\tilde{\alpha}_i\tilde{\alpha}_i}(E) \;
Q_{\tilde{e}\tilde{\mu}}^\dagger(t_\odot) Q_{\tilde{e}\tilde{\tau}}^{\dagger}(t_{\mbox{\tiny R}})
\nonumber
\\&+&\mu
\sum_{i,j=1}^3 \sum_{E^\prime(\neq E)}
\frac{T_{\tilde{\alpha}_i\tilde{\alpha}_j}(E)T_{\tilde{\alpha}_j\tilde{\alpha}_i}(E^\prime)}{\frac{1}{2E}-\frac{1}{2E^\prime}}\msp.
\end{eqnarray}
In order to show that Eq. (\ref{invariants factorized}) is true, one should
substitute the factored form of the operator $Q$ given in Eq.  (\ref{Q tilde})
into Eq. (\ref{3 flavor invariants fb}) and use the definition of the rotated
flavor basis given in Eq.  (\ref{transformation mu-tau}). Note that the
quadratic part of the constants of motion have the same form in the rotated
flavor basis as implied by Eq. (\ref{rotational invariance}).  

\subsection{Effective One Particle Approximation}
\label{subsection: mean field}

The Hilbert space of a self interacting neutrino ensemble grows exponentially
with the number of particles 
so that even with the symmetries described in this paper, diagonalization of the full
many-body Hamiltonian is a formidable task. For this reason one usually resorts to an
\emph{effective one particle} approximation which replaces the system of  
mutually interacting neutrinos with a system of free particles moving in an
average (mean) field. This approach can be formulated with the \emph{operator
product linearization} in which the quadratic term representing mutual interactions of particles
is approximated by 
\begin{subequations}
\begin{equation} 
\label{RPA approximation} 
{\cal O}_1 {\cal O}_2 \sim  
{\cal O}_1 \langle {\cal O}_2 \rangle + \langle {\cal O}_1 \rangle {\cal O}_2 - 
\langle {\cal O}_1 \rangle \langle {\cal O}_2 
\rangle\msp. 
\end{equation} 
Here the expectation values are calculated with respect to a state
$|\Psi\rangle$ which represents the whole system and it is assumed that this state
satisfies the condition 
\begin{equation} 
\label{RPA condition} 
\langle {\cal O}_1  {\cal O}_2 \rangle = \langle {\cal O}_1 \rangle \langle
{\cal O}_2\rangle\msp, 
\end{equation} 
\end{subequations}
so that the expectation values of both sides of Eq. (\ref{RPA approximation})
agree with each other. Usually, the condition in Eq. (\ref{RPA condition}) can
only be satisfied by a restricted class of states in the Hilbert space. In 
Ref. \cite{Balantekin:2006tg} two of us showed that $SU(2)$ or 
$SU(3)$ coherent states can be used for this purpose in the case of two or three
flavors, respectively. 

Application of the operator product linearization to neutrino Hamiltonian given
in Eq. (\ref{vacuum and self hamiltonian}) yields
\begin{equation}
\label{vacuum and self hamiltonian mf}
H_{\mbox{\tiny MF}}=\sum_{E}\sum_{i=1}^3 \frac{\Delta_i^2}{6E}T_{ii}(E)
+\frac{\mu}{2} \sum_{i,j=1}^3 S_{ij}T_{ji}\msp,
\end{equation}
where we define
\begin{equation}
\label{define mean field}
S_{ij}(E,\vec{p}\,)=2\langle T_{ij}(E,\vec{p}\,)\rangle\msp,
\end{equation}
and adopt the same summation conventions for $S_{ij}(E,\vec{p}\,)$ as in Eqs.
(\ref{sumconvention}) and (\ref{sumconvention2}). The factor of $2$ 
in Eq. (\ref{define mean field}) is introduced to account for the fact that when we
linearize a quadratic term as in Eq. (\ref{RPA approximation}), two linear terms
appear on the right hand side.   

Note that the quadratic interaction term that we linearize involve $SU(3)$
generators for which Eq. (\ref{RPA condition}) is only satisfied by $SU(3)$
coherent states \cite{Balantekin:2006tg}. These coherent states involve no
quantum entanglement, i.e., they are in the form of a product of the
one-particle states:
\begin{equation} 
\begin{split}
\label{RPA State} 
|\Psi\rangle\equiv& 
|\psi(\vec{p}_1)\rangle\otimes |\psi(\vec{p}_2)\rangle 
\otimes\dots\otimes|\psi(\vec{p}_N)\rangle\\ 
&\otimes|\bar{\psi}(\vec{p}_1)\rangle\otimes |\bar{\psi}(\vec{p}_2)\rangle 
\otimes\dots\otimes|\bar{\psi}(\vec{p}_N)\rangle. 
\end{split}
\end{equation}
Here the states $|\psi(\vec{p}_k)\rangle$ and
$|\bar{\psi}(\vec{p}_k)\rangle$ represent a single neutrino and antineutrino,
respectively.  They are not necessarily flavor states but can be a superposition
of different flavor or mass eigenstates.  These single particle states are
computed as a function of time by solving a set of \emph{mean field consistency
equations} which guarantee that the mean field evolves in line with the
evolution of the individual particles in the system because all particles
contribute to the mean field.  In order to find these equations, one should
first note that the Heisenberg equation of motion for the operator
$T_{ij}(E,\vec{p}\,)$ is given by
\begin{eqnarray}
\label{eom1}
-i\frac{d}{dt}T_{ij}(E,\vec{p}\,)&=&[H_{\mbox{\tiny MF}},T_{ij}(E,\vec{p}\,)]\\
&=&\frac{\delta m_{ij}^2}{2E}T_{ij}(E,\vec{p}\,)
\nonumber\\
&+&\frac{\mu}{2}\sum_{k=1}^3
\left(S_{ik}T_{kj}(E,\vec{p}\,)-S_{kj}T_{ik}(E,\vec{p}\,)\right)\msp.
\nonumber
\end{eqnarray}
Taking the expectation value of both sides of Eq. (\ref{eom1}) gives 
\begin{eqnarray}
\label{eom2}
-i\frac{d}{dt}S_{ij}(E,\vec{p}\,)
&=&\frac{\delta m_{ij}^2}{2E}S_{ij}(E,\vec{p}\,)\\
&+&\frac{\mu}{2}\sum_{k=1}^3
\left(S_{ik}S_{kj}(E,\vec{p}\,)-S_{kj}S_{ik}(E,\vec{p}\,)\right)\msp,
\nonumber
\end{eqnarray}
which are the mean field consistency equations to be solved to determine 
$S_{ij}(E,\vec{p}\,)$ and hence the state in Eq. (\ref{RPA State}). 

In the mean field approximation, the many-body invariants considered above are
no longer exactly conserved. This is not surprising because when the state of a
particle undergoes a small change as a consequence of its interaction with
another particle, the conservation principle requires the latter to undergo
exactly the opposite change. This requirement obviously cannot be satisfied in a
mean field type approximation \cite{deShalit:1974}. However, the expectation
values of the many-body invariants considered in the previous subsection still
remain constant under the mean field dynamics. In other words, the quantities 
\begin{eqnarray}
\label{3 flavor mean field invariants}
I_E &\equiv& 2\langle h_E \rangle\\
&=&\sum_{i=1}^3\frac{\Delta_i^2}{3} S_{ii}(E) 
+\frac{\mu}{2}\sum_{i,j=1}^3
\sum_{\substack{E^\prime\\ (E^\prime\neq E)}}
\frac{S_{ij}(E)S_{ji}(E^\prime)}{\frac{1}{2E}-\frac{1}{2E^\prime}}
\nonumber
\end{eqnarray}
obey
\begin{equation}
\label{conservation}
\frac{d}{dt}\langle I_E\rangle=0\msp,
\end{equation}
for every $E$. One can easily confirm Eq. (\ref{conservation}) by
taking the derivative of Eq. (\ref{3 flavor mean field invariants})  
and using the mean field equations given in Eq. (\ref{eom2}). 

It is instructive to calculate the values of the constants of motion in the mean
field approximation for neutrinos which are emitted during the cooling phase of
a proto-neutron star after a core collapse supernova explosion.  Initially
$S_{\alpha_i\alpha_j}(E)$ are nonzero only for $i=j$ because all neutrinos are
emitted in flavor states and there is no mixing near the neutron star surface.
Neutrinos reach a thermal equilibrium before they are released from the
proto-neutron star so that the diagonal elements are given by
\begin{equation}
\label{distribution}
\begin{split}
&\langle S_{\alpha_i\alpha_i}(p)\rangle=
\frac{2L}{2\pi R^2 F_3(0)}
\frac{1}{T_{\alpha_i}^4}\frac{p^2}{1+e^{p/T_{\alpha_i}}}\msp,
\\
&\langle S_{\alpha_i\alpha_i}(-p)\rangle=
\frac{-2L}{2\pi R^2 F_3(0)}
\frac{1}{T_{\bar{\alpha}_i}^4}\frac{p^2}{1+e^{p/T_{\bar{\alpha}_i}}}\msp.
\end{split}
\end{equation}
Here $L$ denotes the neutrino luminosity and $R$ denotes the radius of the
neutrino-sphere. We assume that both quantities are the same for all
neutrino and antineutrino flavors.  The Fermi integral $F_3(0)$ corresponding to
zero chemical potential is equal to $7\pi^2/120$.  In Eq. (\ref{distribution}),
the temperature of the $\nu_{\alpha_i}$ and $\bar{\nu}_{\alpha_i}$ are
respectively shown by $T_{\alpha_i}$ and $T_{\bar{\alpha}_i}$. Model independent
arguments tell us that these temperatures obey the hierarchy
\begin{equation}
T_{\nu_e}<T_{\bar{\nu}_e}<T_{\nu_x}=T_{\bar{\nu}_x}\msp,
\end{equation}
where $x=\mu,\tau$. 

Note that near the proto-neutron star, the neutrino luminosity is very large
($L=10^{51}$ ergs/s for the cooling period of the proto-neutron star). In this
case the quadratic terms in the conserved quantities given in Eq. (\ref{3 flavor
mean field invariants}) are at least nine orders of
magnitude larger than the linear terms so that the linear terms can be safely ignored.
As for the quadratic terms in Eq. (\ref{3 flavor mean field invariants}), they
have the same form in both mass and flavor basis as emphasized above (see Eq.
(\ref{3 flavor invariants fb}) and the text that follows it).  
Therefore, the values of the conserved quantities can be obtained by using Eq.
(\ref{distribution}) as follows: 
\begin{widetext}
\begin{equation}
\label{mean field invariants}
\begin{split}
I_p&=I\sum_{i=1}^3 
\frac{1}{T_{\alpha_i}^4}\frac{p^2}{1+e^{p/T_{\alpha_i}}}
\int dq
\left(\frac{\frac{1}{T_{\alpha_i}^4}\frac{q^2}{1+e^{q/T_{\alpha_i}}}}{\frac{1}{2p}-\frac{1}{2q}}
-\frac{\frac{1}{T_{\bar{\alpha}_i}^4}\frac{q^2}{1+e^{q/T_{\bar{\alpha}_i}}}}{\frac{1}{2p}+\frac{1}{2q}}
\right)\msp,\\
I_{-p}&=I \sum_{i=1}^3
\frac{1}{T_{\bar{\alpha}_i}^4}\frac{p^2}{1+e^{p/T_{\bar{\alpha}_i}}}
\int dq 
\left(\frac{\frac{1}{T_{\bar{\alpha}_i}^4}\frac{q^2}{1+e^{q/T_{\bar{\alpha}_i}}}}{\frac{1}{2p}-\frac{1}{2q}}
-\frac{\frac{1}{T_{\alpha_i}^4}\frac{q^2q}{1+e^{q/T_{\alpha_i}}}}{\frac{1}{2p}+\frac{1}{2q}}
\right)\msp.
\end{split}
\end{equation}
\end{widetext}
In writing Eq. (\ref{mean field invariants}), we take the continuum limit 
\begin{equation}
\label{continuum limit}
\frac{1}{V}\sum_{\vec{q}}\to\frac{1}{(2\pi)^3}\int d^3\vec{q}
\end{equation}
and define a common proportionality constant  
\begin{equation}
\label{define A}
I=\frac{RG_F}{\sqrt{2}}\left(\frac{L}{2\pi^2R^2F_3(0)}\right)^2\msp.
\end{equation}
We also take into account that the neutrinos are all going away from the
proto-neutron star so that the angular part of $d\vec{q}$ integrates to
$2\pi$ rather than $4\pi$. 
\begin{figure} 
\begin{center} 
\includegraphics[width=\columnwidth]{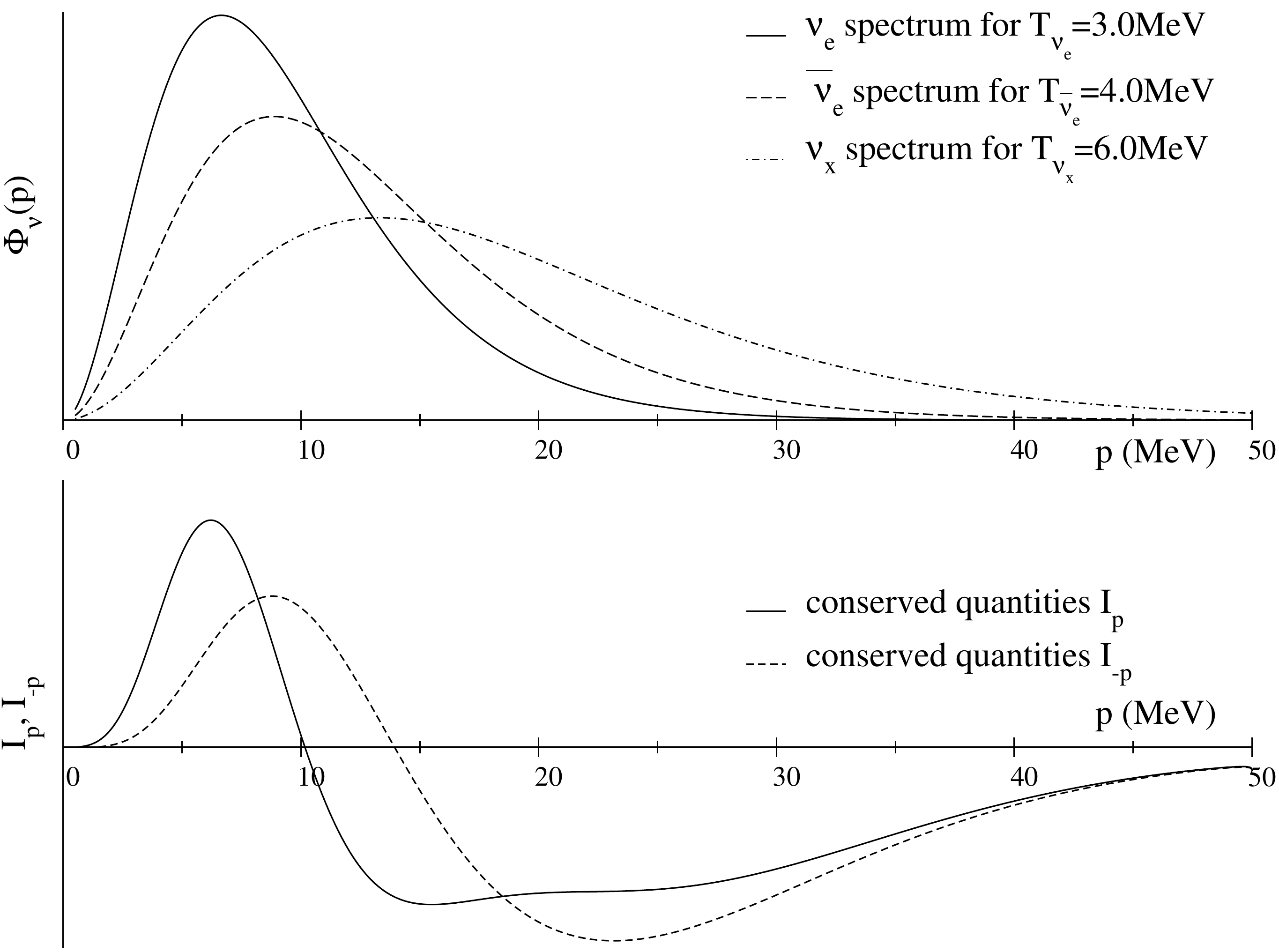} 
\caption{Energy spectrum of the neutrinos emanating from the surface of a
proto-neutron star and the corresponding invariants.  We adopted a
representative set of neutrino temperatures given by $T_{\nu_e}=3.0$,
$T_{\bar{\nu}_e}=4.0$ MeV and $T_{\nu_x}=T_{\bar{\nu}_x}=6.0$ MeV and calculated
the values of the invariants from Eq. (\ref{mean field invariants}).}
\label{conserved fig} 
\end{center} 
\end{figure} 

The values of the invariants calculated from Eq. (\ref{mean field invariants})
for initial neutrino distributions with a representative set of neutrino
temperatures \cite{Yoshida:2006qz, Hayakawa:2010zza, kajino-san-JPhysG}
$T_{\nu_e}=3.0$ MeV, $T_{\bar{\nu}_e}=4.0$ MeV and
$T_{\nu_x}=T_{\bar{\nu}_x}=6.0$ MeV are shown in Figure \ref{conserved fig}.
Note that the values of the invariants depend on the CP-violating Dirac phase
only through the linear term of Eq. (\ref{3 flavor mean field invariants}) which
we ignored in the case of a core collapse supernova (see the discussion above
Eq. (\ref{mean field invariants})).

\section{Magnetic Moment}
\label{section: magnetic moment}

In those astrophysical sources where neutrinos are produced abundantly, it is
also typical to find strong magnetic fields so that even tiny electromagnetic
properties of neutrinos may be consequential. As mentioned in the Introduction,
in the current paradigm of particle physics neutrinos have tiny amounts of
anomalous magnetic moments due to charged particle loops. However, various theories
beyond the Standard Model predict much larger values (see Ref. \cite{Giunti:2014ixa}
and references therein). In this section, we consider the effect of the neutrino
dipole moments on the flavor evolution of neutrinos as they propagate in a
magnetic field. In particular we show that the above mentioned factorization of
CP-violating Dirac phase is not valid under these circumstances, i.e., there is
an interplay between the CP-violating and electromagnetic effects in neutrino
flavor transformation.  

\subsection{Dirac Neutrinos}

Interaction of fermions with a classical electromagnetic field through 
anomalous electric and magnetic dipole moments is described by the Pauli Lagrangian (see
Eq. (91) of Ref.  \cite{Pauli:1941zz} or Section 2-2-3 of Ref.
\cite{Zuber:1980}). In the case of neutrinos, those interactions 
can cause transitions between different types so that the dipole moments should
be represented by matrices, i.e., Pauli Lagrangian is given by
\begin{equation}
\label{pauli}
\mathcal{L}_\mu
=\sum_{i,j=1}^3\bar{\psi}_i\, \frac{1}{2}\mu_{ij} \sigma^{\mu\nu} F_{\mu\nu}\,
\psi_j\msp.
\end{equation}
Here, we use the Greek indices $\mu,\nu=0,1,2,3$ to denote space-time components
and the Latin indices $i,j=1,2,3$ to denote the neutrino mass basis. Summation
convention is adopted for space-time indices but not for the neutrino mass or
flavor indices.  Note that although $\mu_{ij}$ in Eq. (\ref{pauli}) contains
contributions from both electric and magnetic dipole moments of neutrinos, we
follow the convention and refer to it simply as the magnetic moment. In fact,
since the neutrino is ultra relativistic, it \emph{sees} an electric field in
its rest frame and interacts with it through its electric dipole moment even
when there is only a magnetic field present in the environment (see, for
example, appendix E of Ref. \cite{Giunti:2014ixa} for a detailed account).
Note that the hermiticity of the Lagrangian in Eq. (\ref{pauli}) requires that the
magnetic moment is an hermitian matrix, i.e.,
\begin{equation}
\label{hermiticity}
\mu_{ij}=\mu_{ji}^*\msp.
\end{equation}
The dipole moments are defined in the mass basis as indicated in Eq.
(\ref{pauli}) but since neutrinos are produced and detected in flavor states,
physically relevant quantities are effective dipole moments which depend on the
mixing parameters and the energy of the neutrino as well as the distance it
travels from the source.  

In Eq. (\ref{pauli}), $F^{\mu\nu}$ denotes the electromagnetic field tensor and
$\sigma^{\mu\nu}$ is given by
\begin{equation}
\sigma^{\mu\nu}=\frac{i}{2}[\gamma^\mu,\gamma^\nu]\msp.
\end{equation}
We adopt the Euclidean metric $g^{\mu\nu}=\mbox{diag}(1,-1,-1,-1)$ in which case the
interaction term takes the form 
\begin{equation}
\label{sigmaF}
\mu_{ij}\frac{1}{2}\sigma^{\mu\nu}F_{\mu\nu}=\mu_{ij}(i\vec{\alpha}\cdot\vec{E}+\vec{\Sigma}\cdot\vec{B})\msp,
\end{equation}
where 
\begin{equation}
\label{Sigma}
\alpha^k=\gamma^0\gamma^k \qquad \mbox{and} \qquad
\Sigma^k=\gamma^0\gamma^k\gamma^5\msp. 
\end{equation}
In both the Dirac and the chiral representations of the $\gamma$-matrices, $\Sigma^k$
defined in Eq. (\ref{Sigma}) is equal to 
\begin{equation}
\label{Sigma matrix}
\Sigma^k=\begin{pmatrix} \sigma^k & 0 \\ 0 & \sigma^k \end{pmatrix}\msp,
\end{equation}
where $\sigma^k$ are ordinary Pauli matrices. But to be specific, throughout the paper we
use the chiral representation given by
\begin{equation}
\label{gammas}
\gamma^0=\begin{pmatrix} 0 & -I \\ -I & 0 \end{pmatrix},
\qquad
\gamma^k=\begin{pmatrix} 0 & \sigma^k \\ -\sigma^k & 0 \end{pmatrix}.
\end{equation}

One can write down the Hamiltonian density for neutrinos propagating in an
external magnetic field by using Eqs. (\ref{pauli}) and (\ref{sigmaF}) as
\begin{equation}
\label{hamiltonian density}
\mathcal{H}_\mu=
\sum_{i,j=1}^3 \bar{\psi}_i\mu_{ij}\vec{\Sigma}\cdot\vec{B}\psi_j\msp,
\end{equation}
where we set $\vec{E}=0$.
In order to obtain the corresponding many-body Hamiltonian, we
integrate the Hamiltonian density over the space coordinates, 
\begin{equation}
\label{hamiltonian}
H_\mu=\int d^3 \vec{r} \;
\sum_{i,j=1}^3 \bar{\psi}_i\;\mu_{ij}\vec{\Sigma}\cdot\vec{B}\;\psi_j\msp, 
\end{equation}
and use the expansion of the field operator in terms of the plane waves
with definite helicity given by
\begin{eqnarray}
\label{helicity expansion}
&&\psi_i(t,\vec{r})=\int\frac{d^3\vec{p}}{(2\pi)^3}\\
\times&&\sum_{h=\pm}
\left(a_{ih}(\vec{p}\,)u_{h}(\vec{p}\,)e^{-i(pt-\vec{p}\cdot\vec{r})} 
+ b_{ih}^\dagger(\vec{p}\,)v_{h}(\vec{p}\,)e^{i(pt-\vec{p}\cdot\vec{r})} \right)\msp.
\nonumber
\end{eqnarray}
Note that, in this section, we use an integration over the continuous values of
the momentum rather than a sum over discrete values and we no longer use the
convention introduced in Eq. (\ref{convention}).  In Eq. (\ref{helicity
expansion}), $u_h(\vec{p}\,)$ and $v_{-h}(\vec{p}\,)$ are plane wave solutions
for particles and antiparticles, respectively, with helicity $h$. In the ultra
relativistic limit, they are given by  
\begin{equation}
\label{u and v helicity}
\begin{split}
-u_{+}(\vec{p}\,)&=v_{-}(\vec{p}\,)=\begin{pmatrix} \chi^{(+)} \\ 0 \end{pmatrix}\msp,
\\
u_{-}(\vec{p}\,)&=- v_{+}(\vec{p}\,)=\begin{pmatrix} 0 \\ \chi^{(-)} \end{pmatrix}\msp,
\end{split}
\end{equation}
where $\chi^{(h)}$ are the helicity eigenstates which are given as follows:
\begin{equation}
\label{helicity eigenstates}
\chi^{(+)}=
\begin{pmatrix}
\displaystyle{\cos\frac{\theta}{2}} \\ e^{i\phi}\displaystyle{\sin\frac{\theta}{2}}
\end{pmatrix}\msp,
\quad
\chi^{(-)}=
\begin{pmatrix}
-e^{-i\phi}\displaystyle{\sin\frac{\theta}{2}} \\ \displaystyle{\cos\frac{\theta}{2}}
\end{pmatrix}\msp.
\end{equation}
In Eq. (\ref{helicity eigenstates}), $\theta$ and $\phi$ denote polar and
azimuthal angles of the momentum $\vec{p}$, respectively. Note that according to
Eqs. (\ref{helicity expansion}) and (\ref{u and v helicity}), the operators
$a_{ih}(p)$ annihilate neutrinos in the $i^{\mbox{\tiny th}}$ mass eigenstate
with helicity $h$ whereas the operators $b^\dagger_{ih}(p)$ create antineutrinos
in the $i^{\mbox{\tiny th}}$ mass eigenstate with helicity $-h$.   

Substituting the expansion of the field operator given in Eq. (\ref{helicity
expansion}) into the interaction Hamiltonian in Eq. (\ref{hamiltonian}),
assuming that $\vec{B}$ is a uniform field, and keeping only those terms which
are relevant to the propagation of neutrinos in the limit where $E\gg m$ yields
the following result for the Hamiltonian:
\begin{eqnarray}
\label{appendix hamiltonian helicity}
H_\mu\!=\!
\int\!\! d^3\vec{p}\! \!\! \sum_{h,h^\prime=\pm} \!\!\!
&& \mu_{ij}\!\!
\left\{ a_{ih}^\dagger(\vec{p}\,)[\bar{u}_h(\vec{p}\,)
\;\vec{\Sigma}\cdot\vec{B}\;u_{h^\prime}(\vec{p}\,)]a_{jh^\prime}(\vec{p}\,)
\right.  \nonumber \\
&&\mkern-25mu\left.+
b_{ih}(\vec{p}\,)[\bar{v}_h(\vec{p}\,)
\;\vec{\Sigma}\cdot\vec{B}\;v_{h^\prime}(\vec{p}\,)]b_{jh^\prime}^\dagger(\vec{p}\,)
\right\}\msp.
\end{eqnarray}
The expressions which appear in square brackets in Eq. (\ref{appendix
hamiltonian helicity})  can be easily calculated using Eqs. (\ref{Sigma matrix})
and (\ref{u and v helicity}). The result is given by 
\begin{equation}
\label{terms}
\begin{split}
\bar{u}_h(\vec{p}\,)\;\vec{\Sigma}\;u_{h^\prime}(\vec{p}\,)
&= \left(\hat{n}_\theta+ih\hat{n}_\phi\right) \delta_{h^\prime,-h}\msp,
\\
\bar{v}_h(\vec{p}\,)\;\vec{\Sigma}\;v_{h^\prime}(\vec{p}\,)
&= 
-\left(\hat{n}_\theta+ih\hat{n}_\phi\right) \delta_{h^\prime,-h}\msp.
\end{split}
\end{equation}
Here $\hat{n}_\theta$ and $\hat{n}_\phi$ are two unit vectors which are
orthogonal to the direction of motion of the neutrino. In other words,
$\hat{p}=\vec{p}/|\vec{p}|$, $\hat{n}_\theta$ and $\hat{n}_\phi$
form an orthonormal basis for the spherical coordinates in the momentum space.
In terms of the Cartesian unit vectors they are given by
\begin{eqnarray}
\label{spherical units}
\hat{p} &=& \sin\theta\cos\phi\:\hat{x} + \sin\theta\sin\phi\:\hat{y} +
\cos\theta\:\hat{z} \msp,
\nonumber\\
\hat{n}_\theta &= &\cos\theta\cos\phi\:\hat{x} + \cos\theta\sin\phi\:\hat{y} -
\sin\theta\:\hat{z} \msp,
\\
\hat{n}_\phi &=& \sin\phi\:\hat{x} - \cos\phi\:\hat{y}\msp.
\nonumber
\end{eqnarray}
Substitution of the results in Eq.  (\ref{terms}) into Eq. (\ref{appendix
hamiltonian helicity}) yields the following result:
\begin{eqnarray}
\label{Hmag Dirac}
H_\mu\!=\!\int\!\! &d^3& \!\vec{p}
\sum_{i,j=1}^3\mu_{ij}B_\perp\left(a_{i+}^\dagger(\vec{p}\,) a_{j-}(\vec{p}\,) +  b_{j+}^\dagger(\vec{p}\,)b_{i-}(\vec{p}\,) \right) 
\nonumber\\
&+& \mbox{h.c.} 
\end{eqnarray}
Here, $B_\perp=(\hat{n}_\theta+i\hat{n}_\phi)\cdot\vec{B}$ denotes the component
of the magnetic field which is perpendicular to the direction of neutrino
propagation. One can always rotate the plane perpendicular to the direction of
neutrino propagation to make $\hat{n}_\phi\cdot\vec{B}=0$ so that $B_\perp$ can
be assumed to be real.  

In order to express the flavor evolution of neutrinos in the presence of a
strong magnetic field, one should write the Hamiltonian given in Eq. (\ref{Hmag
Dirac}) in flavor basis. The transformation of left handed neutrinos from mass
to flavor basis is discussed in Section \ref{section: transformations}. However,
the right handed Dirac neutrinos do not take part in weak interactions so the
choice of \emph{flavor} basis for them is completely arbitrary. For our
purposes, this choice is of no practical consequences and we simply leave the
right handed neutrinos in mass basis in our formulas.

The Hamiltonian in Eq. (\ref{Hmag Dirac}) can be expressed in the flavor basis
by using the inverse of Eq. (\ref{transformation}):   
\begin{eqnarray}
H_\mu=Q\int d^3 \vec{p}
\sum_{i,j=1}^3\mu_{ij}B_\perp&&\left(a_{i+}^\dagger(\vec{p}\,) a_{\alpha_j-}(\vec{p}\,)\right. \\
&&\mkern-25mu+
\left.  b_{j+}^\dagger(\vec{p}\,)b_{\alpha_i-}(\vec{p}\,)
\right)Q^\dagger 
+\mbox{h.c.} 
\nonumber
\end{eqnarray}
The form of the transformation operator given in Eq. (\ref{Q tilde}) is once
again useful in examining the dependence of this Hamiltonian on CP-violating
Dirac phase. As was the case in Section \ref{subsection: Neutrino
Propagation with CP Violation}, the rightmost $Q_{\mu\tau}$ in Eq. (\ref{Q tilde}) transforms   
the left handed neutrino degrees of freedom into the rotated flavor basis
leading to 
\begin{widetext}
\begin{equation}
\label{Hmag Dirac in flavor}
H_\mu=
S_{\tilde{\tau}}^\dagger Q_{\tilde{e}\tilde{\tau}}(t_{\mbox{\tiny R}})
Q_{\tilde{e}\tilde{\mu}}(t_{\odot})S_{\tilde{\tau}}
\int d^3 \vec{p}\sum_{i,j=1}^3
\;\mu_{ij}B_\perp\left(a_{i+}^\dagger(\vec{p}\,) a_{\tilde{\alpha}_j-}(\vec{p}\,) +
b_{j+}^\dagger(\vec{p}\,)b_{\tilde{\alpha}_i-}(\vec{p}\,)
\right)
S_{\tilde{\tau}}^\dagger Q_{\tilde{e}\tilde{\mu}}^\dagger(t_{\odot})
Q_{\tilde{e}\tilde{\tau}}^\dagger(t_{\mbox{\tiny R}}) S_{\tilde{\tau}}
+ \mbox{h.c.} 
\end{equation}

Unlike the case in the vacuum oscillations, the operator $S_{\tilde{\tau}}$
which contains the CP-violating phase does not commute with the terms in
parenthesis so, strictly speaking, we cannot disentangle the CP-violating effects
from those of the magnetic moment. However, one can show that 
\begin{equation}
\label{S transformation}
S_{\tilde{\tau}}a_{\tilde{\alpha}_i-}S_{\tilde{\tau}}^\dagger 
=\sum_{j=1}^3S_{ij} a_{\tilde{\alpha}_j-}
\quad \mbox{and} \quad
S_{\tilde{\tau}}b_{\tilde{\alpha}_i-}S_{\tilde{\tau}}^\dagger 
=\sum_{j=1}^3S_{ij}^* b_{\tilde{\alpha}_j-}
\end{equation}
are satisfied where $S_{ij}$ is given by
\begin{equation}
\label{S matrix}
S=\begin{pmatrix}
1 & 0 & 0\\ 0 & 1 & 0 \\ 0 & 0 & e^{i\delta}
\end{pmatrix}\msp.
\end{equation}
As a result, one can define an \emph{effective} magnetic moment $\mu^{\mbox{\tiny eff}}$
as 
\begin{equation}
\label{effective mu dirac}
\mu^{\mbox{\tiny eff}}=\mu S=
\begin{pmatrix}
\mu_{11} & \mu_{12} & \mu_{13} e^{i\delta}\\
\mu_{12}^* & \mu_{22} & \mu_{23}e^{i\delta}\\
\mu_{13}^* & \mu_{23}^* & \mu_{33}e^{i\delta} 
\end{pmatrix}
\end{equation}
and write the Hamiltonian in Eq. (\ref{Hmag Dirac in flavor}) as 
\begin{equation}
\label{Hmag with effective mu}
H_\mu=
S_{\tilde{\tau}}^\dagger \left( Q_{\tilde{e}\tilde{\tau}}^{(0)}Q_{\tilde{e}\tilde{\mu}}
\int d^3 \vec{p}
\;B_\perp\sum_{i,j=1}^3\left(\mu^{\mbox{\tiny eff}}_{ij}a_{i+}^\dagger(\vec{p}\,) a_{\tilde{\alpha}_j-}(\vec{p}\,) +
{\mu^{\mbox{\tiny eff}}_{ij}}^*b_{j+}^\dagger(\vec{p}\,)b_{\tilde{\alpha}_i-}(\vec{p}\,)
\right)
Q_{\tilde{e}\tilde{\mu}}^\dagger
{Q_{\tilde{e}\tilde{\tau}}^{(0)}}^\dagger
+ \mbox{h.c.} 
\right)
S_{\tilde{\tau}}\msp.
\end{equation}
\end{widetext}

This tells us that the Hamiltonian describing neutrinos in a strong magnetic
field can be factorized as 
\begin{equation} 
H_\mu=S_{\tilde{\tau}}^\dagger \tilde{H}_{\mu^{\mbox{\tiny eff}}}S_{\tilde{\tau}}\msp, 
\end{equation}
where $S_{\tilde{\tau}}$ contains the CP-violating Dirac phase and is given by
Eq. (\ref{define S}). The Hamiltonian $\tilde{H}_{\mu^{\mbox{\tiny eff}}}$ is
given by the expression in parenthesis in Eq. (\ref{Hmag with effective mu}).
It describes neutrinos with an effective magnetic moment in the rotated flavor
basis (as indicated by the tilde sign) and does not contain the CP-violating
Dirac phase \emph{explicitly}. 
However, the effective magnetic moment defined in Eq. (\ref{effective mu
dirac}) is not a unitary matrix. The appearance of $\mu^{\mbox{\tiny eff}}$ for
neutrinos and ${\mu^{\mbox{\tiny eff}}}^*$ for antineutrinos in Eq.  (\ref{Hmag
with effective mu}) reflects the CP violation.  This clearly shows that the
effects of CP violation and magnetic moment are intertwined and cannot be
separated. 

But aside from proving this point, the formulation developed in this section can
also be practical. For example, one can consider the neutrino propagation in the
presence of a matter background and self interactions as well as a magnetic
field by using the Hamiltonian (see Eq. (\ref{total hamiltonian factorized compact}))
\begin{equation}
\label{total hamiltonian with effective mu}
H=S_{\tilde{\tau}}^\dagger \left(H_{\tilde{\mbox{\footnotesize v}}}^{(0)}+H_{\tilde{\mbox{\footnotesize m}}}
+H_{\tilde{\mbox{\footnotesize s}}}+\tilde{H}_{\mu^{\mbox{\tiny
eff}}}\right)S_{\tilde{\tau}}\msp.
\end{equation} 
The term in the
parenthesis in Eq. (\ref{total hamiltonian with effective mu}) includes CP
violation only implicitly through $\mu^{\mbox{\tiny eff}}$ which, in most cases, can be simply
studied to the first order in perturbation theory. In such a
calculation, CP-violating phase will appear only linearly and create a
minimal complication.  The full effect of the CP-violating phase can later be
included using Eq.  (\ref{U}).

\subsection{Majorana Neutrinos}

If the neutrinos are of Majorana type, then the part of the Lagrangian in Eq.
(\ref{pauli}) involving the symmetric component of $\mu_{ij}$ vanishes
automatically once the Majorana condition $\psi_i^c=\psi_i$ is imposed.
Therefore, the magnetic moment can be taken as an antisymmetric matrix for Majorana
neutrinos: 
\begin{equation}
\label{antisymmetry}
\mu_{ij}=\mu_{ji}^* 
\quad\mbox{and}\quad
\mu_{ij}=-\mu_{ji}\msp.
\end{equation}
This tells us that, for the Majorana neutrinos, the diagonal magnetic moments
vanish and the non-diagonal ones are purely imaginary.  Also note that, once we
impose the Majorana condition, the Lagrangian in Eq. (\ref{pauli}) should  be
divided by $2$ in order to avoid double counting of neutrino and antineutrino
degrees of freedom.  In our notation introduced in Eqs. (\ref{helicity
expansion}), (\ref{u and v helicity}) and (\ref{helicity eigenstates}), the
Majorana condition amounts to 
\begin{equation}
\label{majorana condition}
b_{i+}(\vec{p}\,) = a_{i-}(\vec{p}\,) 
\quad\mbox{and}\quad
b_{i-}(\vec{p}\,) =- a_{i+}(\vec{p}\,)\msp. 
\end{equation}
Another important point is the fact that, although neutrinos and antineutrinos
are identical implied by Eq. (\ref{majorana condition}), it is conventional to call
Majorana neutrinos with positive helicity \emph{antineutrinos} because, as far
as the production and detection of neutrinos are concerned, the difference
between positive helicity Majorana neutrinos and positive helicity Dirac
antineutrinos is suppressed by neutrino mass/energy. Therefore, for Majorana
neutrinos, we adopt the notation
\begin{equation}
\label{majorana notation}
a_{i-}(\vec{p}\,) = a_i(\vec{p}\,)
\quad\mbox{and}\quad
a_{i+}(\vec{p}\,)= b_i(\vec{p}\,)\msp.
\end{equation}

\begin{widetext}
Substituting Eqs. (\ref{antisymmetry}), (\ref{majorana condition}) and
(\ref{majorana notation}) in Eq. (\ref{Hmag Dirac}) and dividing it by $2$
yields the corresponding Hamiltonian for Majorana neutrinos:  
\begin{equation}
\label{Hmag Majorana}
H_\mu=\int d^3 \vec{p} \sum_{i,j=1}^3\mu_{ij}B_\perp b_{i}^\dagger(\vec{p}\,) a_{j}(\vec{p}\,) \; +\; \mbox{h.c.} 
\end{equation}

Unlike the case in Dirac neutrinos, the transformation of Majorana \emph{antineutrinos}
from mass to flavor basis is fixed by Eq. (\ref{transformation}). Together with
Eq. (\ref{Q tilde}), this leads to
\begin{equation}
H_\mu=
S_{\tilde{\tau}}^\dagger Q_{\tilde{e}\tilde{\tau}}(t_{\mbox{\tiny R}})
Q_{\tilde{e}\tilde{\mu}}(t_{\odot})S_{\tilde{\tau}}
\left(\int d^3 \vec{p} \sum_{i,j=1}^3
\mu_{ij}B_\perp b_{\tilde{\alpha}_i}^\dagger(\vec{p}\,) a_{\tilde{\alpha}_j}(\vec{p}\,)
+ \mbox{h.c.} \right)
S_{\tilde{\tau}}^\dagger Q_{\tilde{e}\tilde{\mu}}^\dagger(t_{\odot})
Q_{\tilde{e}\tilde{\tau}}^\dagger(t_{\mbox{\tiny R}}) S_{\tilde{\tau}}
\msp.
\end{equation}
Using Eqs. (\ref{S transformation}) and (\ref{S matrix}) which are still valid in the Majorana case, one 
obtains
\begin{equation}
H_\mu=
S_{\tilde{\tau}}^\dagger Q_{\tilde{e}\tilde{\tau}}(t_{\mbox{\tiny R}})
Q_{\tilde{e}\tilde{\mu}}(t_{\odot})
\left(\int d^3 \vec{p} \sum_{i,j=1}^3
\mu^{\mbox{\tiny eff}}_{ij}B_\perp b_{\tilde{\alpha}_i}^\dagger(\vec{p}\,) a_{\tilde{\alpha}_j}(\vec{p}\,)
+ \mbox{h.c.} \right)
Q_{\tilde{e}\tilde{\mu}}^\dagger(t_{\odot})
Q_{\tilde{e}\tilde{\tau}}^\dagger(t_{\mbox{\tiny R}}) S_{\tilde{\tau}}
\msp,
\end{equation}
\end{widetext}
where $\mu^{\mbox{\tiny eff}}$ is defined as follows
\begin{equation}
\label{effective mu majorana}
\mu^{\mbox{\tiny eff}}=S\mu S=
\begin{pmatrix}
0 & \mu_{12} & \mu_{13} e^{i\delta}\\
-\mu_{12} & 0 & \mu_{23}e^{i\delta}\\
-\mu_{13} e^{i\delta} & -\mu_{23}e^{i\delta} & 0 
\end{pmatrix}\msp.
\end{equation}

As is the case in Dirac neutrinos, the effective magnetic moment is not an
hermitian matrix but it is still antisymmetric. We see that Eq. (\ref{total
hamiltonian with effective mu})  and the comments following that equation are
also valid for Majorana neutrinos provided that the effective magnetic moment is
now given by Eq. (\ref{effective mu majorana}). 

\section{Summary and Conclusions}

In this paper, we considered the flavor evolution of neutrinos which are subject
to refractive effects due to both self interactions and matter background. We
attempted to a comprehensive study of the problem by taking into account its full
many-body nature in the three flavor mixing scenario with the effects of
possible CP violation and anomalous magnetic moment included. Since our
perspective was exclusively based on the symmetries of the problem,
important environmental details were left out of our analysis, such as a
specific core collapse supernova model for matter and magnetic field profiles.

We showed that, in its exact many-body formulation, the system exhibits several
dynamical symmetries in such a way that one has a constant of motion for each
allowed neutrino and antineutrino energy mode. We expressed these constants of
motion in terms of the generators of the $SU(3)$ flavor transformations. In the
case of the effective one particle approximation, we showed that the expectation
values of these constants of motion remain invariant under the mean field
dynamics. The dynamical symmetries considered in this paper are valid under a
set of \emph{ideal} conditions, i.e., when the single angle approximation is
adopted, the \emph{net} electron background is negligible, and the volume occupied by
the neutrinos is fixed ($\mu=$ constant). We also showed that these dynamical
symmetries are not broken even when CP symmetry is violated in neutrino
oscillations. 

Even away from the ideal conditions mentioned above, the constants of motion
presented in this paper can still be useful by providing a convenient set of
variables to work with because one can always decompose the Hamiltonian into an
\emph{ideal} and a \emph{non-ideal} part as 
\begin{equation}
H=H_{\mbox{\footnotesize ideal}}+H_{\mbox{\footnotesize non-ideal}}\msp,
\end{equation}
such that, although the \emph{constants of motion} will now evolve in time, their
evolution will only be due to the non-ideal part, i.e.,  
\begin{equation}
-i\frac{d}{dt}h_E=[H_{\mbox{\footnotesize non-ideal}},h_E]
\end{equation}
since they commute with the ideal part of the Hamiltonian. 

In this paper, we also showed that the CP violation effects factor out of the Hamiltonian and
the evolution operator not only in the effective one particle picture adopted by
the mean field type approximations, but also in the full many-body picture. This
conclusion is exact as long as the neutrino magnetic moment is not considered
but even when one includes the neutrino dipole moments into the analysis,
CP violation can still be studied independently as long as an effective magnetic
moment is defined which includes the Dirac CP-violating phase in an implicit
way. Clearly, the effects due to CP violation and magnetic moment are
intertwined in an inseparable way even in this formulation because the
definition of the effective magnetic moment is different for neutrinos and
antineutrinos. However, such a formulation is still useful because it allows us to
include the CP-violating effects in a seamless and methodical way into
analytical and numerical calculations. On the practical side, even when the
neutrino magnetic moment is not ignored, this formulation locks
the CP-violating phase only into the magnetic moment which is very small and can
be conveniently studied only to the first order in a perturbation approach.

\vspace*{3mm}
\noindent 
Y.P. is grateful to University of Wisconsin for their hospitality where part of
this work was completed and to American Physical Society for the International
Travel Grant Award which allowed her visit. Y.P. and A.B.B. thank National
Astronomical Observatory of Japan  for their hospitality. We also thank to CETUP*
2013 organizers for allowing a stimulating environment for critical discussion
of some of our results.  This work was supported 
in part by the Scientific and Technological Research Council of Turkey
(T{\"{U}}B{\.{I}}TAK) under project number 112T952,
in part by the U.S. National Science Foundation Grant No.
PHY-1205024, in part by the University of Wisconsin Research Committee with funds
granted by the Wisconsin Alumni Research Foundation, 
and in part by Grants-in-Aid for Scientific Research of JSPS (26105517, 24340060) of the
Ministry of Education, Culture, Sports, Science and Technology of Japan.

\appendix 
\section{Reduction to two flavor scheme} 
\label{Appendix reduction} 

The constants of motion given in Eq. (\ref{3 flavor invariants mb}) for three
mixing flavors reduce to those that were presented earlier in Ref.
\cite{Pehlivan:2011hp} in the context of a two flavor mixing scheme.  In order
to show this, we first consider the two neutrino \emph{isospin} operators 
\begin{equation}
\begin{split}
\label{isospin redifine}
J^+(p,\vec{p}\,) &= T_{12}(p,\vec{p}\,)\msp,  
\qquad
J^-(p,\vec{p}\,) = T_{21}(p,\vec{p}\,)\msp, \\ 
&J^0(p,\vec{p}\,) = \frac{T_{11}(p,\vec{p}\,)-T_{22}(p,\vec{p}\,)}{2}\msp,\\ 
\end{split}
\end{equation}
which are similar to Eqs. (\ref{isospin2}) and (\ref{isospin2anti}) except that
the negative energy formulation for antineutrinos is now incorporated.
We adopt the same summation convention for these isospin operators as in
Eqs. (\ref{sumconvention}) and (\ref{sumconvention2}). Note that we 
choose to work with the first two mass eigenstates but this choice is completely
arbitrary. It is easy to show that 
\begin{equation}
\label{reduction main}
\sum_{i,j=1}^2T_{ij}(E)T_{ji}(E^\prime)=2\vec{J}(E)\cdot\vec{J}(E^\prime)
+\frac{1}{2}N_{12}(E)N_{12}(E^\prime)\msp,
\end{equation}
where $N_{12}(E)=T_{11}(E)+T_{22}(E)$ is the total number of neutrinos
($E>0$) or antineutrinos ($E>0$) ) in the first two mass eigenstates with energy $E$.

Next, we consider the Hamiltonian given in Eq. (\ref{vacuum and self
hamiltonian}) but restrict the range of the sums over the mass eigenstates that
appear in this Hamiltonian to the first two mass eigenstates only. Note that
there is no need to set $m_3=0$, i.e., the result is independent of the value of
$m_3$.  Then, using the definitions given in Eq. (\ref{isospin redifine})
together with Eq. (\ref{reduction main}), leads to
\begin{equation}
\label{reduced Hamiltonian}
H_{\mbox{\tiny two flavors}}= \sum_{E}\frac{\delta m_{12}^2}{2E}J^0(E)
+\mu \vec{J}\cdot\vec{J}\msp.
\end{equation}
In deriving Eq. (\ref{reduced Hamiltonian}),  we discarded some terms which are
proportional to $N_{12}(E)$ because  it commutes with the rest of the
Hamiltonian and is proportional to identity. 

The constants of motion given in Eq. (\ref{3 flavor invariants mb}) can
similarly be reduced to the two flavor mixing scheme in a similar way.
Restricting the sums over the mass eigenstates to this first two mass
eigenstates only, using   Eqs. (\ref{isospin redifine}) and (\ref{reduction
main}), and dropping the terms proportional to $N_{12}(E)$ leads to
\begin{equation} 
\label{Invariants} 
h_E=\delta m_{12}^2J^0(E)+2\mu\sum_{E^\prime (\neq E)}
\frac{\vec{J}(E)\cdot\vec{J}(E^\prime)}{\frac{1}{2E}-\frac{1}{2E^\prime}}\msp. 
\end{equation} 
Dividing Eq. (\ref{Invariants}) by $\delta m_{12}^2$ gives the
same many-body invariants which were presented in Ref. \cite{Pehlivan:2011hp}.


\vfill

\begin{thebibliography}{99}

\bibitem{Lesgourgues:2012uu} 
  J.~Lesgourgues and S.~Pastor,
  Adv.\ High Energy Phys.\  {\bf 2012}, 608515 (2012)
  [arXiv:1212.6154 [hep-ph]].



\bibitem{Lesgourgues:2014zoa} 
  J.~Lesgourgues and S.~Pastor,
  arXiv:1404.1740 [hep-ph].



\bibitem{Hannestad:2006zg} 
  S.~Hannestad,
  Ann.\ Rev.\ Nucl.\ Part.\ Sci.\  {\bf 56}, 137 (2006)
  [hep-ph/0602058].



\bibitem{Dolgov:2002wy} 
  A.~D.~Dolgov,
  Phys.\ Rept.\  {\bf 370}, 333 (2002)
  [hep-ph/0202122].



\bibitem{Burrows:1990ts} 
  A.~Burrows,
  Ann.\ Rev.\ Nucl.\ Part.\ Sci.\  {\bf 40}, 181 (1990).



\bibitem{Kotake:2005zn} 
  K.~Kotake, K.~Sato and K.~Takahashi,
  Rept.\ Prog.\ Phys.\  {\bf 69}, 971 (2006)
  [astro-ph/0509456].



\bibitem{Beacom:2010kk} 
  J.~F.~Beacom,
  Ann.\ Rev.\ Nucl.\ Part.\ Sci.\  {\bf 60}, 439 (2010)
  [arXiv:1004.3311 [astro-ph.HE]].


\bibitem{Mathews:2014qba} 
  G.~J.~Mathews, J.~Hidaka, T.~Kajino and J.~Suzuki,
  Astrophys.\ J.,\ in press (2014) [arXiv:1405.0458 [astro-ph.CO]].



\bibitem{Raffelt:1996}
    Georg G.~Raffelt,
    \emph{Stars as Laboratories for Fundamental Physics},
    University of Chicago Press,
    1996.



\bibitem{Narayan:2001qi} 
  R.~Narayan, T.~Piran and P.~Kumar,
  astro-ph/0103360.



\bibitem{Ruffert:1998qg} 
  M.~Ruffert and H.~T.~Janka,
  Astron.\ Astrophys.\  {\bf 344}, 573 (1999)
  [astro-ph/9809280].



\bibitem{Popham:1998ab} 
  R.~Popham, S.~E.~Woosley and C.~Fryer,
  Astrophys.\ J.\  {\bf 518}, 356 (1999)
  [astro-ph/9807028].



\bibitem{Matteo:2002ck} 
  T.~D.~Matteo, R.~Perna and R.~Narayan,
  Astrophys.\ J.\  {\bf 579}, 706 (2002)
  [astro-ph/0207319].



\bibitem{Chen:2006rra} 
  W.~-X.~Chen and A.~M.~Beloborodov,
  Astrophys.\ J.\  {\bf 657}, 383 (2007)
  [astro-ph/0607145].


\bibitem{Malkus:2012ts} 
  A.~Malkus, J.~P.~Kneller, G.~C.~McLaughlin and R.~Surman,
  Phys.\ Rev.\ D {\bf 86}, 085015 (2012)
  [arXiv:1207.6648 [hep-ph]].


\bibitem{fuller_mayle}
  G.~M.~Fuller, R.~W.~Mayle, J.~R.~Wilson and D.~N.~Schramm,
  Astrophys.\ J.\  {\bf 322}, 795 (1987).



\bibitem{Pantaleone:1992eq} 
  J.~T.~Pantaleone,
  Phys.\ Lett.\ B {\bf 287}, 128 (1992).



\bibitem{Pantaleone:1992xh} 
  J.~T.~Pantaleone,
  Phys.\ Rev.\ D {\bf 46}, 510 (1992).



\bibitem{Sawyer:2005jk} 
  R.~F.~Sawyer,
  Phys.\ Rev.\ D {\bf 72}, 045003 (2005)
  [hep-ph/0503013].



\bibitem{Sigl:1992fn} 
  G.~Sigl and G.~Raffelt,
  Nucl.\ Phys.\ B {\bf 406}, 423 (1993).



\bibitem{Friedland:2006ke} 
  A.~Friedland, B.~H.~J.~McKellar and I.~Okuniewicz,
  Phys.\ Rev.\ D {\bf 73}, 093002 (2006)
  [hep-ph/0602016].



\bibitem{Friedland:2003eh} 
  A.~Friedland and C.~Lunardini,
  JHEP {\bf 0310}, 043 (2003)
  [hep-ph/0307140].



\bibitem{Friedland:2003dv} 
  A.~Friedland and C.~Lunardini,
  Phys.\ Rev.\ D {\bf 68}, 013007 (2003)
  [hep-ph/0304055].


\bibitem{Balantekin:2006tg} 
  A.~B.~Balantekin and Y.~Pehlivan,
  J.\ Phys.\ G {\bf 34}, 47 (2007)
  [astro-ph/0607527].



\bibitem{Duan:2010bg} 
  H.~Duan, G.~M.~Fuller and Y.~-Z.~Qian,
  Ann.\ Rev.\ Nucl.\ Part.\ Sci.\  {\bf 60}, 569 (2010)
  [arXiv:1001.2799 [hep-ph]].
  
  

\bibitem{Kostelecky:1994dt} 
  V.~A.~Kostelecky and S.~Samuel,
  Phys.\ Rev.\ D {\bf 52}, 621 (1995)
  [hep-ph/9506262].



\bibitem{Samuel:1995ri} 
  S.~Samuel,
  Phys.\ Rev.\ D {\bf 53}, 5382 (1996)
  [hep-ph/9604341].



\bibitem{Duan:2005cp} 
  H.~Duan, G.~M.~Fuller and Y.~-Z.~Qian,
  Phys.\ Rev.\ D {\bf 74}, 123004 (2006)
  [astro-ph/0511275].



\bibitem{Duan:2006an} 
  H.~Duan, G.~M.~Fuller, J. Carlson and Y.~-Z.~Qian,
  Phys.\ Rev.\ D {\bf 74}, 105014 (2006)
  [astro-ph/0606616]; 
  G.~G.~Raffelt and A.~Y.~.Smirnov,
  Phys.\ Rev.\ D {\bf 76}, 081301 (2007)
  [Erratum-ibid.\ D {\bf 77}, 029903 (2008)]
  [arXiv:0705.1830 [hep-ph]]; 
  Phys.\ Rev.\ D {\bf 76}, 125008 (2007)
  [arXiv:0709.4641 [hep-ph]].



\bibitem{Raffelt:2011yb} 
  G.~G.~Raffelt,
  Phys.\ Rev.\ D {\bf 83}, 105022 (2011)
  [arXiv:1103.2891 [hep-ph]].



\bibitem{Pehlivan:2011hp} 
  Y.~Pehlivan, A.~B.~Balantekin, T.~Kajino and T.~Yoshida,
  Phys.\ Rev.\ D {\bf 84}, 065008 (2011)
  [arXiv:1105.1182 [astro-ph.CO]].



\bibitem{Lee:1977tib} 
  B.~W.~Lee and R.~E.~Shrock,
  Phys.\ Rev.\ D {\bf 16}, 1444 (1977).



\bibitem{Marciano:1977wx} 
  W.~J.~Marciano and A.~I.~Sanda,
  Phys.\ Lett.\ B {\bf 67}, 303 (1977).



\bibitem{Balantekin:2013sda}
  A.~B.~Balantekin and N.~Vassh,
Phys. Rev. D {\bf 89}, 073013 (2014)
  [arXiv:1312.6858 [hep-ph]].



\bibitem{Beda:2013mta} 
  A.~G.~Beda, V.~B.~Brudanin, V.~G.~Egorov, D.~V.~Medvedev, V.~S.~Pogosov, E.~A.~Shevchik, M.~V.~Shirchenko and A.~S.~Starostin {\it et al.},
  Phys.\ Part.\ Nucl.\ Lett.\  {\bf 10}, 139 (2013).



\bibitem{Raffelt:1999gv} 
  G.~G.~Raffelt,
  Phys.\ Rept.\  {\bf 320}, 319 (1999).


\bibitem{Viaux:2013lha} 
  N.~Viaux, M.~Catelan, P.~B.~Stetson, G.~G.~Raffelt, J.~Redondo, A.~A.~R.~Valcarce and A.~Weiss,
  Phys.\ Rev.\ Lett.\  {\bf 111}, 231301 (2013)
  [arXiv:1311.1669 [astro-ph.SR]].



\bibitem{Giunti:2014ixa} 
  C.~Giunti and A.~Studenikin,
  arXiv:1403.6344 [hep-ph].



\bibitem{Lim:1987tk}
  C.~-S.~Lim and W.~J.~Marciano,
  Phys.\ Rev.\ D {\bf 37}, 1368 (1988);
  E.~K.~Akhmedov,
  Sov.\ J.\ Nucl.\ Phys.\  {\bf 48}, 382 (1988)
  [Yad.\ Fiz.\  {\bf 48}, 599 (1988)];
  A.~B.~Balantekin, P.~J.~Hatchell and F.~Loreti,
  Phys.\ Rev.\ D {\bf 41}, 3583 (1990).



\bibitem{deGouvea:2012hg} 
  A.~de Gouvea and S.~Shalgar,
  JCAP {\bf 1210}, 027 (2012)
  [arXiv:1207.0516 [astro-ph.HE]].



\bibitem{deGouvea:2013zp} 
  A.~de Gouvea and S.~Shalgar,
  JCAP {\bf 1304}, 018 (2013)
  [arXiv:1301.5637 [astro-ph.HE]].



\bibitem{An:2013zwz}
  F.~P.~An {\it et al.}  [Daya Bay Collaboration],
  Phys.\ Rev.\ Lett.\  {\bf 112}, 061801 (2014)
  [arXiv:1310.6732 [hep-ex]];
  F.~P.~An {\it et al.}  [Daya Bay Collaboration],
  Chin.\ Phys.\ C {\bf 37}, 011001 (2013)
  [arXiv:1210.6327 [hep-ex]];
  F.~P.~An {\it et al.}  [DAYA-BAY Collaboration],
  Phys.\ Rev.\ Lett.\  {\bf 108}, 171803 (2012)
  [arXiv:1203.1669 [hep-ex]].



\bibitem{Ahn:2012nd}
  J.~K.~Ahn {\it et al.}  [RENO Collaboration],
  Phys.\ Rev.\ Lett.\  {\bf 108}, 191802 (2012)
  [arXiv:1204.0626 [hep-ex]].



\bibitem{Abe:2012tg}
  Y.~Abe {\it et al.}  [Double Chooz Collaboration],
  Phys.\ Rev.\ D {\bf 86}, 052008 (2012)
  [arXiv:1207.6632 [hep-ex]].



\bibitem{Adams:2013qkq} 
  C.~Adams {\it et al.}  [LBNE Collaboration],
  arXiv:1307.7335 [hep-ex].


\bibitem{::2013kaa} 
  S.~K.~Agarwalla {\it et al.}  [LAGUNA-LBNO Collaboration],
  arXiv:1312.6520 [hep-ph].



\bibitem{Balantekin:2007es} 
  A.~B.~Balantekin, J.~Gava and C.~Volpe,
  Phys.\ Lett.\ B {\bf 662}, 396 (2008)
  [arXiv:0710.3112 [astro-ph]].



\bibitem{Kneller:2009vd} 
  J.~P.~Kneller and G.~C.~McLaughlin,
  Phys.\ Rev.\ D {\bf 80}, 053002 (2009)
  [arXiv:0904.3823 [hep-ph]].



\bibitem{Gava:2010kz} 
  J.~Gava and C.~Volpe,
  Nucl.\ Phys.\ B {\bf 837}, 50 (2010)
  [arXiv:1002.0981 [hep-ph]].



\bibitem{Gava:2008rp} 
  J.~Gava and C.~Volpe,
  Phys.\ Rev.\ D {\bf 78}, 083007 (2008)
  [arXiv:0807.3418 [astro-ph]].



\bibitem{Duan:2007fw} 
  H.~Duan, G.~M.~Fuller and Y.~-Z.~Qian,
  Phys.\ Rev.\ D {\bf 76}, 085013 (2007)
  [arXiv:0706.4293 [astro-ph]].



\bibitem{Duan:2007bt} 
  H.~Duan, G.~M.~Fuller, J.~Carlson and Y.~Z.~Qian, 
  Phys.\ Rev.\ Lett.\  {\bf 99}, 241802 (2007)
  [arXiv:0707.0290 [astro-ph]].



\bibitem{Wolfenstein:1977ue} 
  L.~Wolfenstein,
  Phys.\ Rev.\ D {\bf 17}, 2369 (1978).



\bibitem{Mikheev:1986wj} 
  S.~P.~Mikheev and A.~Y.~.Smirnov,
  Nuovo Cim.\ C {\bf 9}, 17 (1986).



\bibitem{Mikheev:1986gs} 
  S.~P.~Mikheev and A.~Y.~.Smirnov,
  Sov.\ J.\ Nucl.\ Phys.\  {\bf 42}, 913 (1985)
  [Yad.\ Fiz.\  {\bf 42}, 1441 (1985)].



\bibitem{Bardeen:1957mv} 
  J.~Bardeen, L.~N.~Cooper and J.~R.~Schrieffer, 
  Phys.\ Rev.\  {\bf 108}, 1175 (1957). 



\bibitem{Gaudin1} 
  M.~Gaudin,  
  J. Physique \textbf{37}(1976), 1087. 


 
\bibitem{Gaudin2} 
  M.~Gaudin,  
  Collection du Commissariat a l'\'{e}nergie atomique, Masson, Paris, 1983. 


   
\bibitem{Cambiaggio} 
  M.~C.~Cambiaggio, A.~M.~F.~Rivas and M.~Saraceno, 
  Nucl. Phys. \textbf{A624}(1997) 157 
  [arXiv:nucl-th/9708031]. 



\bibitem{yuzb} 
 A.~A.~Yuzbashyan, B.~L.~Altshuler, V.~B.~Kuznetsov, and V.~E.~Enolskii, J.
 Phys. A: Math. Gen. {\bf 38}, 7831 (2005) [arXiv:cond-mat/0407501].   



\bibitem{yuzb2}
  E.~A.~Yuzbashyan, 
  Phys. Rev. B 78, 184507 (2008)
  [arXiv:0807.3181 [cond-mat.supr-con]].



\bibitem{deShalit:1974}
    A.~De Shalit and H.~Feshbach,
    \emph{Theoretical Nuclear Physics: Nuclear Structure v. 1},
    John Wiley \& Sons Inc, 1974 (Chapter IV.23).



\bibitem{Yoshida:2006qz} 
  T.~Yoshida, T.~Kajino, H.~Yokomakura, K.~Kimura, A.~Takamura and D.~H.~Hartmann,
  Phys.\ Rev.\ Lett.\  {\bf 96}, 091101 (2006)
  [astro-ph/0602195].



\bibitem{Hayakawa:2010zza} 
  T.~Hayakawa, T.~Kajino, S.~Chiba and G.~J.~Mathews,
  Phys.\ Rev.\ C {\bf 81}, 052801 (2010)
  [arXiv:1012.5700 [astro-ph.HE]].



\bibitem{kajino-san-JPhysG}
T.\ Suzuki and T.\ Kajino, J.\ Phys.\ G: Nucl.\ Part.\ Phys.\ {\bf 40} 
083101 (2013). 



\bibitem{Pauli:1941zz} 
  W.~Pauli,
  Rev.\ Mod.\ Phys.\  {\bf 13}, 203 (1941).


\bibitem{Zuber:1980}
    C.~Itzykson and J.~B.~Zuber
    \emph{Quantum Field Theory},
    McGraw-Hill, 1980.


\end{thebibliography}
\end{document}